\documentclass[%
 aip,
 amsmath,amssymb,
 reprint,%
]{revtex4-1}

\usepackage{graphicx}
\usepackage{dcolumn}
\usepackage{bm}

\usepackage[utf8]{inputenc}
\usepackage[T1]{fontenc}
\usepackage[english]{babel}
\usepackage{mathptmx}
\usepackage{etoolbox}

\usepackage{booktabs}
\usepackage{tabularx}
\usepackage{xcolor}
\usepackage{longtable}
\usepackage{hyperref}
\usepackage[caption=false]{subfig}

\usepackage{etoolbox}
\patchcmd{\thebibliography}{\selectlanguage{english}}{}{}{}

\newcommand{\lex}[1]{$\rm{#1}$} 

\newcommand{\N}{\lex{N} }
\newcommand{\Ntw}{\lex{N_2} }
\newcommand{\NtwA}{\lex{N_2(A)} }
\newcommand{\NtwB}{\lex{N_2(B)} }
\newcommand{\NtwC}{\lex{N_2(C)} }
\newcommand{\Ntwap}{\lex{N_2(a')} }
\newcommand{\Ox}{\lex{O} }
\newcommand{\Otw}{\lex{O_2} }
\newcommand{\Oxd}{\lex{O(D)}}
\newcommand{\Oth}{\lex{O_3} }
\newcommand{\NO}{\lex{NO} }
\newcommand{\Np}{\lex{N^+} }
\newcommand{\Ntwp}{\lex{N_2^+} }
\newcommand{\Nthp}{\lex{N_3^+} }
\newcommand{\Nforp}{\lex{N_4^+} }
\newcommand{\Otwp}{\lex{O_2^+} }
\newcommand{\Op}{\lex{O^+} }
\newcommand{\Oforp}{\lex{O_4^+} }
\newcommand{\NOp}{\lex{NO^+} }
\newcommand{\Om}{\lex{O^-} }
\newcommand{\Otwm}{\lex{O_2^-} }
\newcommand{\Nxd}{\lex{N(D)}}
\newcommand{\Nxp}{\lex{N(P)}}
\newcommand{\ele}{\lex{e^-} }
\newcommand{\NtOt}{\lex{N_2-O_2} }

\newcommand{\etal}{\textit{et al.}}

\newcommand{\allothers}{\cite{obrusnik_electric_2018,pancheshnyi_zdplaskin_2008,troe_temperature_2005,ilyin_emission_2022,park_rate_2008, niu_assessment_2018, kim_modification_2021,bakReducedSetAir2015,poggieNumericalSimulationNanosecondpulse2012,campbell_recombination_1997,bodrovEffectElectricField2013,florescu-mitchell_dissociative_2006,SIGLODatabase,cicman_rate_2003} }

\makeatletter
\def\@email#1#2{%
 \endgroup
 \patchcmd{\titleblock@produce}
  {\frontmatter@RRAPformat}
  {\frontmatter@RRAPformat{\produce@RRAP{*#1\href{mailto:#2}{#2}}}\frontmatter@RRAPformat}
  {}{}
}%
\makeatother
\begin{document}

\preprint{AIP/123-QED}

\title[]{Characteristics of Femtosecond Laser Induced Filament and Energy Coupling by Nanosecond Laser Pulse in Air}
\author{S. Pokharel}
\author{A. A. Tropina}%

\email{pokharel\_sagar@tamu.edu}
\affiliation{ 
Department of Aerospace Engineering, Texas A\&M University, TX 77843
}%

\date{\today}

\begin{abstract}

This study presents a detailed plasma kinetics model for laser-induced non-equilibrium plasmas in atmospheric pressure air, incorporating a self-consistent energy balance and refined rate expressions within a three-temperature framework. The model is validated against experimental data of femtosecond-laser-induced filaments, showing good agreement in electron dynamics and gas temperature. The analysis focuses on femtosecond filament decay kinetics and characteristic properties across varying initial electron densities and electron temperatures, including cases with oxygen addition and its influence on decay behavior. The study further examines energy coupling between the femtosecond filament and nanosecond laser pulse, identifying dominant kinetic pathways and optimal time delays through a comparative analysis of single-pulse and dual-pulse plasmas. Additionally, results indicate that temporal shaping of the nanosecond laser pulse intensity enhances dual-pulse performance relative to Gaussian pulses.

\end{abstract}

\maketitle

\section{Introduction}

Laser energy deposition and laser generated plasmas have numerous potential applications, such as aerodynamic flow control \cite{limbach_characterization_2015}, fuel-air mixture ignition \cite{Tropina2020Ign}, flame stabilization \cite{Dumitrache2017ControlIgnition}, directed energy deposition \cite{Dumitrache2016LaserApproach},  hazardous gas tracing \cite{Shneider2011TailoringPulse}, machining and fabrication for modern semiconductors \cite{Barkusky2007}, waveguiding \cite{jhajj_demonstration_2014}, laser diagnostics \cite{edwards_simultaneous_2015}, etc.
Depending on the characteristics of the plasma kernel, e.g., electron number density, electron energy, size, shape, translational and vibrational temperatures, and pressure, different flow dynamics are observed at hydrodynamic time scales $t > 1 \mu s$. For efficient development of applications based on the controlled laser energy deposition, a detailed understanding of plasma kinetics, hydrodynamics, and associated aero-optical effects is required. 

The exponential nature of plasma buildup while using nanosecond (ns) laser pulses often results in uncontrolled energy deposition. To address this, a dual-pulse approach, incorporating both ns-ns and femtosecond (fs)-ns configurations, has been proposed \cite{limbach_characterization_2015,dumitrache_control_2017,yalinLaserPlasmaFormation2014}. This method enables the decoupling of initial ionization and seeded electron generation by the first pulse from the subsequent energy deposition, allowing for greater control over plasma formation.
le energy addition by the second heating pulse \cite{Shneider2011TailoringPulse,Tropina2017,tropina_plasma_2021-1}.  Dual-laser pulse setups, particularly those incorporating fs filaments, are being explored for controlled plasma generation and energy deposition. Fs filaments are characterized by high-intensity pre-ionization and self-focusing, and facilitate long-distance propagation and waveguiding of subsequent laser pulses. These systems, enhanced by additional energy sources, enable rapid triggering of pulsed discharges and advanced diagnostics for low-temperature plasmas and ignition. The efficiency of energy deposition in optimal pulse pairings is influenced by the interaction between the second laser pulse and the initial plasma kernel. Thus, a comprehensive characterization of fs filaments at atmospheric pressure, along with their coupling with ns laser pulses, is crucial for applications in aerospace and other advanced technological fields.

The properties of laser-induced plasmas in ambient air vary significantly due to factors such as laser pulse energy, system jitter, focusing configurations, humidity, and inhomogeneities caused by particulates. While existing studies have mainly focused on initial electron density using interferometry \cite{aleksandrov_decay_2016,sun_determination_2011,chizhov_interferometry_2015}, plasma kinetics and other properties under highly non-equilibrium conditions remain underexplored. Experimental techniques face challenges caused by fast time scales and small filament dimensions. Recent experimental investigations have utilized techniques such as Thomson scattering \cite{TwodimensionalHighResolution2023a}, microwave scattering \cite{bakLaserIntensityShaping2024,papeer_temporal_2013}, and optical emission spectroscopy \cite{blanchardCharacterizationFemtosecondLaser2024} for more comprehensive studies of femtosecond-laser-induced filament dynamics. However, studies on their coupling with nanosecond laser pulses \cite{papeerExtendedLifetimeHigh2014,bakLaserIntensityShaping2024, limbach_characterization_2015,Shneider2011TailoringPulse} remain limited.

Théberge et al. \cite{thebergePlasmaDensityFemtosecond2006} demonstrated experimentally and numerically that the plasma density and diameter of the plasma column within the core of a single filament are strongly influenced by the focal length of the lens used to focus the beam, while being only minimally affected by the laser power. The electron density in femtosecond (fs)-laser induced filaments has been reported to range from \( 1.0 \times 10^{14} \, \text{cm}^{-3} \) to \( 1.0 \times 10^{18} \, \text{cm}^{-3} \), with free propagation resulting in the lowest densities and tighter focusing optics yielding higher densities \cite{thebergePlasmaDensityFemtosecond2006,reyesTransitionLinearNonlinearfocusing2018}. For extremely tight focusing, where the numerical aperture (NA) exceeds 0.08, plasma densities up to \( 1.0 \times 10^{19} \, \text{cm}^{-3} \) have been observed \cite{liuTightlyFocusedFemtosecond2010,kiranFocalDynamicsMultiple2010}. Braun et al. \cite{braunSelfchannelingHighpeakpowerFemtosecond1995} demonstrated the self-channeling property of fs laser filaments in air over distances exceeding 20 m. More recent studies with chirped pulses reported propagation distances up to 1 km, retaining 80\% of the initial pulse energy \cite{mechainLongrangeSelfchannelingInfrared2004}.

Quantitative measurements of the temporal evolution of plasma properties in fs laser-induced (FLI) filaments have been reported, including gas temperature \cite{edwards_simultaneous_2015}, vibrational temperature\cite{blanchardCharacterizationFemtosecondPlasma2025}, and electron density \cite{aleksandrov_decay_2016,chizhov_interferometry_2015,papeer_temporal_2013,sun_determination_2011}. Two-dimensional filament properties in argon have been studied by Bak \cite{bakTwodimensionalHighResolution2024}. Recent work utilizing Thomson scattering has experimentally measured electron density, electron temperature, gas temperature, and gas density \cite{urdanetaTemporallyResolvedProperties2025} on the same filament. The influence of the ns-heating pulse on fs laser filaments has been explored in several studies. Papeer demonstrated the slowing down of plasma decay with the heating pulse \cite{papeerExtendedLifetimeHigh2014}, while Polynkin observed the appearance of so-called plasma bullets with the energy addition from the ns-heating pulse \cite{polynkinSeededOpticalBreakdown2012a}. Limbach investigated the absorption of a second pulse beyond breakdown conditions by adding a high energy ns pulse \cite{limbach_characterization_2015}. Energy addition to fs-filaments, which causes  triggering of air-gap discharge \cite{kosareva_remote_2021} and combined fs and microwave discharges \cite{michaelLocalizedMicrowavePlasma2011} were also studied.

Studies related to the fs laser propagation in the media have explored various aspects of laser-induced plasma dynamics, including electron yield calculations and modeling filament conditions for varying power and focusing setups \cite{fengPropagationFemtosecondLaser2022,thebergePlasmaDensityFemtosecond2006}. In the most common approach the electron density is evaluated  considering the clamping intensity of the laser, calculated from the balance of the photoionization yield \cite{perelomov_ionization_1966} and the Kerr focusing, without considering the spatially expanding and contracting structure of the fs filament \cite{peters_considerations_2019, pokharelSelfconsistentModelNumerical2023a}.

For plasma kinetics in laser-induced highly non-equilibrium plasmas, limited studies have validated electron dynamics and thermalization through the comparison with experimental measurements. Aleksandrov developed a reduced reaction scheme fitted to the measured plasma decay curves \cite{aleksandrov_decay_2016}. Papeer \etal \cite{papeer_temporal_2013} employed a detailed plasma kinetics model to fit experimental decay curves and determine quantitative electron density values, but the specifics of the reaction scheme used were not clearly described. The re-energization capabilities required for modeling the coupling of a ns-heating pulse with a fs-filament are often neglected. Modeling plasma kinetics for decay alone is insufficient for dual-pulse laser plasma applications, as the kinetics of plasma decay and re-energization are significantly different \cite{kruger_nonequilibrium_2002}. Shneider investigated the energy addition to fs-filaments in air using a dual laser pulse approach, incorporating detailed plasma kinetics \cite{shneider_tailoring_2011}. The drift-diffusion model has been effectively applied to describe laser plasma dynamics \cite{pokharelSelfconsistentModelNumerical2023a} in nitrogen, showing good agreement with experimental results. The development of plasma kinetics reaction scheme for non-equilibrium plasma in air  often relies on the work of Kossyi \cite{kossyi_kinetic_1992} and other contributions from the literature \cite{popov_associative_2009, shneider_population_2011, obrusnik_electric_2018}.

Accurate modeling of laser-induced plasma phenomena over extended timescales requires self-consistent approaches that integrate multi-species kinetics, energy exchange mechanisms across various energy modes, and dimensional considerations. Significant gaps remain in the characterization of fs-ns dual-laser pulse plasmas and the development of reliable plasma dynamics solvers with robust validation for such applications. Comprehensive models are required to accurately represent these phenomena, focusing on multi-laser interactions (e.g., dual-laser pulses) rather than solely capturing electron decay. Detailed plasma kinetics models are crucial for low-temperature, highly non-equilibrium plasmas below breakdown and subsequent laser-plasma interactions, as they must capture thermochemical non-equilibrium characteristics, plasma revival, and consistent energy thermalization.

This work focuses on the development of comprehensive plasma kinetics models for laser-induced plasmas in atmospheric pressure air, incorporating a self-consistent energy balance for non-equilibrium plasmas and improved rate expressions under a multi-temperature framework. Emphasis is placed on fs-filament dynamics and their coupling with ns laser pulses, addressing challenges in modeling filament behavior, comparing experimental results, and understanding interaction mechanisms. The study includes the development and validation of plasma kinetics models for $N_2-O_2$ mixtures, followed by detailed investigations into fs-filament behavior, dual-pulse plasma interactions, and recent advances in coupling mechanisms and plasma dynamics.

\section{Mathematical Model and Solver Development}

The femtosecond laser-induced plasma channel is distinct from other plasma sources due to its pronounced thermochemical non-equilibrium characteristics. While some applications, such as ignition, aerodynamic flow control, and flow-field modification, rely on thermal effects, others, including sensitive optical diagnostics, precise energy deposition, the prevention of uncontrolled breakdown, and wave-guiding, operate within low-temperature, highly non-equilibrium plasma conditions, necessitating specialized models. Ignition can still be achieved with higher efficiency without realizing such high temperatures reached in thermal post-breakdown plasmas, but necessitates control over the energy deposition process.

In our previous work \cite{pokharelSelfconsistentModelNumerical2023a}, we introduced a drift-diffusion-based fluid model for non-equilibrium plasmas, incorporating detailed chemical kinetics in multiple dimensions. Additionally, we developed a three-temperature plasma kinetic model coupled with a Navier-Stokes solver to accurately simulate the decay of laser induced plasma. The plasma solver, LOTASFOAM \cite{pokharelSelfconsistentModelNumerical2023a}, utilizing OpenFOAM libraries \cite{weller_tensorial_1998}, integrates a multi-temperature plasma model alongside a hybrid central solver for hydrodynamics with wide range of mach numbers. Addressing challenges associated with modeling the thermochemical non-equilibrium nature of laser-induced low-temperature plasma, we identified and resolved several key issues. Our approach focused on achieving a self-consistent balance of energy in plasma reactions, ensuring the inclusion of energy exchange between various energy modes within a reaction. This was crucial for maintaining a consistent energy balance in the plasma, especially over extended timescales, a feature currently lacking in multi-dimensional models. Additionally, our model incorporated appropriate considerations for charge transport involving negative ions and electrons, introducing a separate equation for the electric field. Furthermore, vibrational non-equilibrium was addressed by including a dedicated vibrational energy equation. These enhancements contribute significantly to a more comprehensive and accurate description of laser-induced low-temperature plasma behavior in multi-dimensional models, with extensive validation and robustness demonstrated for nitrogen plasmas at atmospheric pressure \cite{pokharelSelfconsistentModelNumerical2023a,urdanetaImplementationLaserThomson2024,pokharelRefractiveIndexModification2024}.

\begin{figure}
    \centering
    \includegraphics[width=\linewidth]{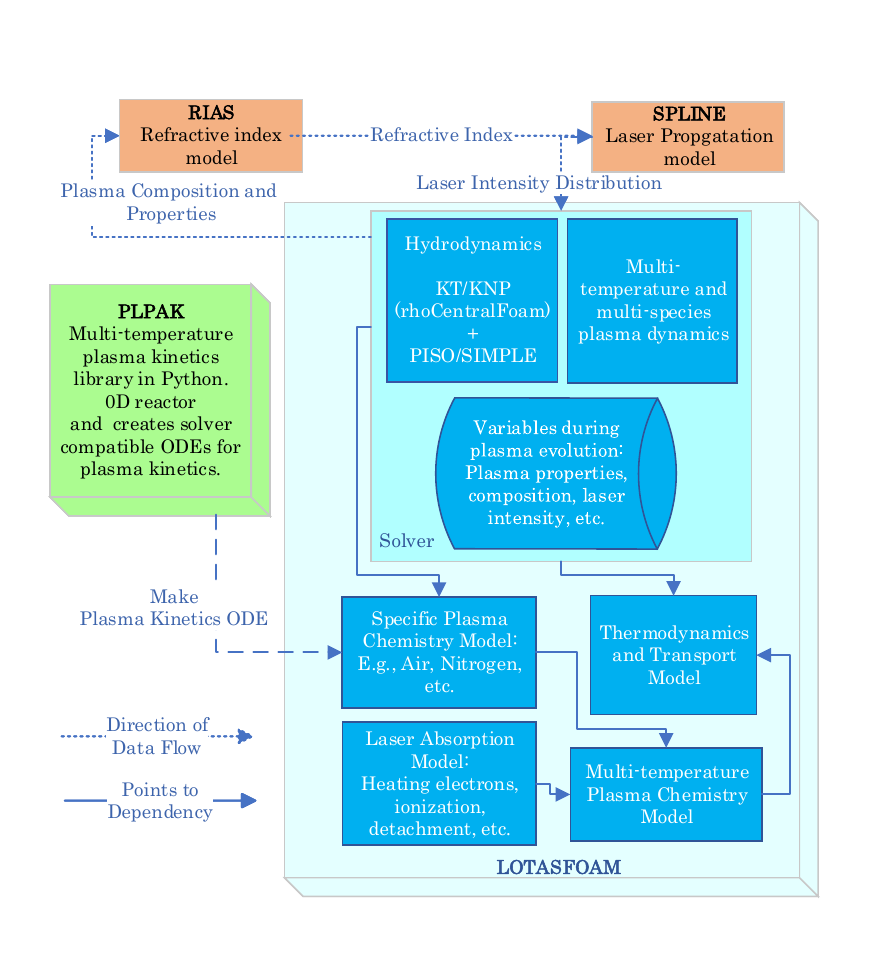}
    \caption{Schematic diagram for the mulit-physics solver network.}
    \label{fig:schemSolverNetwork}
\end{figure}

The plasma dynamics solver is part of a holistic framework that incorporates multi-physics details for studying laser-induced plasmas and their interactions across various applications. Although the presented results do not involve coupling between laser propagation and plasma dynamics, this framework integrates key components, including the SPLINE model for laser propagation, the PLPAK (Package for Low Temperature Plasma Kinetics) library for plasma kinetics development, and the RIAS module for refractive index and laser absorption calculations, all interfaced with the LOTASFOAM solver to form a comprehensive multi-physics solver suit. The schematic of this solver architecture is presented in Figure \ref{fig:schemSolverNetwork}. The detailed mathematical model utilized is explained in a previous publication \cite{pokharelSelfconsistentModelNumerical2023a}, while the extension and implementation of the plasma kinetics for air and additional details related to the coupling of laser-plasma interactions are presented herein.

A significant extension in this work is the development of self-consistent plasma kinetics models for air plasmas, facilitated by the PLPAK library. PLPAK is a standalone utility for developing 0D-3T (zero-dimensional and three-temperature) plasma kinetics models, enabling rapid testing and analysis of reaction mechanisms. It generates plasma-system ODEs for drift-diffusion solvers in major programming languages. The library incorporates BOLSIG+ rates, supports multi temperature-dependent rate expressions, and ensures accurate representation of thermodynamic properties and multi-temperature energy exchange. Key features of PLPAK include ensuring a conservative energy balance through an enthalpy-based formulation for heat of reaction and consistently incorporating fast gas heating, super-elastic collision heating of electrons from vibrational relaxation, and other energy exchange modes. PLPAK with the developed plasma-kinetics reaction mechanism sets for $\mathrm{N_2-O_2}$ mixture is made publicly available to the scientific community and can be accessed at \url{https://github.com/ptroyen/PLPAK}.


\subsection{Plasma Kinetics Model}

\subsubsection{Reaction Mechanism for \NtOt}

To describe the chemical kinetics associated with plasma decay, we extend the plasma kinetics model for pure $N_2$ based on references \cite{peters_considerations_2019,shneider_population_2011,zhang_enhancement_2016}, incorporating accurate and self-consistent energy exchange in chemical reactions and between different modes to ensure energy conservation. The three-temperature plasma kinetics model for \Ntw was validated against experimental measurements of electron density decay for a range of initial electron densities in our earlier work \cite{pokharelSelfconsistentModelNumerical2023a}. The earlier work \cite{pokharelSelfconsistentModelNumerical2023a,peters_considerations_2019} did not consider the distinction of electronically excited species of atomic nitrogen, which is used to explain the slow decay of electrons in nitrogen plasma by Popov \cite{popov_associative_2009} through associative ionization. Therefore, additional new species are included in the earlier model \cite{peters_considerations_2019} to include associative ionization \cite{popov_associative_2009, leonov_femtosecond_2012, volynets_n2_2018}, and further addition of reactions and improvements are made to develop a complete reaction set for a \NtOt plasma system. Thus, we include a total of 24 species: \N, \Ntw, \NtwA, \NtwB, \NtwC, \Ntwap, \Ox, \Otw, \Oxd, \Oth, \NO, \Np, \Ntwp, \Nthp, \Nforp, \Otwp, \Op, \Oforp, \NOp, \Om, \Otwm, \Nxd, \Nxp, and \ele.

The development of the reaction mechanism set is primarily based on the works of Kyossi\cite{kossyi_kinetic_1992}, Shneider\cite{Shneider2011}, and Popov\cite{popov_associative_2009,popovFastGasHeating2011}. Additional reactions or necessary modifications from available literature data \allothers are incorporated into reaction rates to address inconsistencies in previous works, and elucidated as needed. The application of a second pulse can significantly elevate plasma temperatures, necessitating consideration of thermal dissociation, ionization, and atomic recombination, commonly required in weakly ionized plasma modeling in hypersonic flows. For such reactions, Park's \cite{park_rate_2008} air-plasma reaction models are applied. Long-term electron decay is predominantly governed by ion-ion recombination due to Kyossi \cite{kossyi_kinetic_1992}. Electron impact reactions are determined using BOLSIG+ \cite{hagelaar_solving_2005} to compute the electron energy distribution function, utilizing collision cross-section data available from the LaxCat database \cite{SIGLODatabase}.

Nitrogen atoms recombine to form electronically excited states of molecular nitrogen, as described by Popov \cite{popovFastGasHeating2011} and Kyossi \cite{kossyi_kinetic_1992} in the reaction \N + \N + M $\rightarrow$ \lex{N_2(A,B)} + M. However, the reported rates for atomic recombination can be ambiguous when only a single value is provided without considering electronic excited states. Recent works \cite{obrusnik_electric_2018,pancheshnyi_zdplaskin_2008} independently present recombination rates for these processes, enabling differentiation between different electronic states. The electron impact dissociation rates from electronically excited states are sourced from Park \cite{park_rate_2008}. Notably, recombination rates for atomic nitrogen derived from both Park and modified Park models (refer to \cite{kim_modification_2021}, Fig. 7) demonstrate tendencies to either significantly underestimate or overestimate at low temperatures (300 K). Conversely, the rates proposed by Campbell \cite{campbell_recombination_1997} for low-temperature plasma exhibit an overestimation at higher temperatures ($T > 2000$ K). Hence, a combination of both sets of rates, adjusted for temperature, is utilized for atomic nitrogen recombination (R194). The formation of \Nforp clusters is typically considered a three-body process. However, Troe \cite{troe_temperature_2005} demonstrated that it saturates into a two-body process, $\mathrm{N_2^+ + N_2 \rightarrow N_4^+}$, at atmospheric pressure. Consequently, reaction R46 is formulated as a three-body process that saturates to a two-body process at 1 atm, with rates taken from \cite{ilyin_emission_2022}. The \Ntw-plasma kinetics model \cite{peters_considerations_2019,pokharelSelfconsistentModelNumerical2023a}, upon which this work builds, lacked consideration for the distinction of electronically excited species of atomic nitrogen. Popov \cite{popov_associative_2009, leonov_femtosecond_2012, volynets_n2_2018} emphasizes the significance of these species in elucidating the slow decay of electrons in nitrogen plasma via associative ionization. To incorporate this pivotal process, which potentially holds relevance in dual-pulse laser energy deposition scenarios, where the second pulse augments the heating and coupling with the pre-ionized plasma channel, additional new species, namely \Nxd, \Nxp, and \Oxd, are integrated to account for associative ionization.

In the self-consistent plasma kinetics model, inconsistencies in rate expressions can lead to nonphysical changes in associated temperatures. This is because the heat of reactions relies on the difference between reactants and products' enthalpy, thus inconsistent rates contribute to additional errors in the energy equation. Such undesirable outcomes for specific reactions prompt the investigation and refinement of rate expressions, suggesting the inclusion of temperature dependence for such reactions. An example of this arises in the dissociative attachment of \Oth (R83). While rates from Kyossi \cite{kossyi_kinetic_1992} are provided as constants, reactions R83 (endothermic) and R84 (exothermic) involve considerable energy exchange, warranting improved rates and/or expressions. Rates with dependence on electron temperature (\lex{T_e}) for R83-84 are provided by Cicman \cite{cicman_rate_2003} and are used instead. Utilizing rates from Cicman \cite{cicman_rate_2003}, in reaction R83, for instance, a decrease in electron temperature results in a corresponding decrease in the rate, leading to stable electron temperature decay. Bodrov \cite{bodrovEffectElectricField2013} suggested a three-body recombination mechanism for \Otwp, indicating that it proceeds via dissociation to yield atomic oxygen. Hence, the rates for this process are obtained from Bodrov \cite{bodrovEffectElectricField2013}. Consequently, the reaction R72  is depicted as 2\ele + \Otwp $\rightarrow$ \ele + 2\Ox, rather than 2\ele + \Otwp $\rightarrow$ \ele + \Otw.

\subsubsection{ Thermodynamics and Transport\label{sec:ThermoTransport}}

Thermodynamic properties are obtained from the NASA database \cite{noauthor_explosion_nodate}, with enthalpy adjustments for electronically and vibrationally excited species using data from NIST spectroscopic resources \cite{noauthor_nist_nodate} and partition functions from Therm4NEC \cite{hazenberg_consistent_2023}. Mixture-averaged bulk gas transport properties, including mass diffusion coefficient, thermal conductivity, and viscosity, are determined using the mass-averaged mixing rule. Additionally, the mobility of species is required for multi-dimensional models. We utilize BOLSIG+ \cite{hagelaar_solving_2005} along with appropriate cross-sections \cite{noauthor_siglo_nodate} to determine electron mobility and collision frequencies. Experimental measurements take precedence for other ions when available; otherwise, the Langevin model is employed (Eq. \ref{eq:mobility}). Here, $\alpha$ represents the polarizability of the mixture taken as $\alpha = 1.73 \text{\AA}^3$ for air \cite{bohringer_mobilities_1987}, and $M$ is the reduced mass of the ion-neutral pair in atomic mass units. Data for other ions are shown in Table \ref{tab:ion_mobilities} \cite{bohringer_mobilities_1987, ellis_transport_1976}.

\begin{equation}
    \mu_0 = \frac{13.876}{(\alpha M )^{0.5}} ~ \text{cm}^2 \text{V}^{-1} \text{s}^{-1}; ~~~~        \mu = \mu_0 \frac{760}{p_{\text{torr}}} \frac{T_{\text{K}}}{273.16} 
    \label{eq:mobility}
\end{equation}

\begin{table}
    \caption{Reduced mobilities for ions in air plasma.}
    \begin{ruledtabular}
    \begin{tabular}{lclc}
        \textbf{Ions} & \textbf{$\mu_0$ [$\text{cm}^2 \text{V}^{-1} \text{s}^{-1}$]} & \textbf{Ions} & \textbf{$\mu_0$ [$\text{cm}^2 \text{V}^{-1} \text{s}^{-1}$]} \\
        \hline
        $N^+$ & 3.433 & $O_2^+$ & 2.51 \\
        $N_2^+$ & 2.795 & $O_4^+$ & 2.362 \\
        $N_3^+$ & 2.548 & $O_2^-$ & 2.53 \\
        $N_4^+$ & 2.414 & $O^-$ & 3.32 \\
        $O^+$ & 3.32 & $NO^+$ & 2.65 \\
    \end{tabular}
    \end{ruledtabular}
    \label{tab:ion_mobilities}
\end{table}

\subsubsection{ Energy Exchange }

\begin{subequations} \label{eq:Qs}
\begin{align}
Q_{ET} &= \frac{3}{2} n_e k_B (T_e - T) (\nu_{en} + \nu_{ei}) \frac{2 m_e}{m_h} \\
Q_{EV} &= \sum_{l=1}^{l_{\text{max}}} U_l K_l n_v n_e ; \quad Q_{VE} = \sum_{l=1}^{l_{\text{max}}} U_l K_l^{\text{inv}} n_{vl} n_e \\
Q_{VT } &= \frac{(E_v - E_v^0)}{\tau_{VT}}; ~~~ E_v(T_v) =  \frac{n_{v} k_B \theta_v}{\exp{\left( \frac{\theta_v}{T_v} \right)} - 1};
\end{align}
\end{subequations}

For elementary chemical reactions, heating values are derived from the enthalpy difference between reactants and products. Accurate thermodynamic property calculations, particularly species' enthalpy, are crucial within a multi-temperature framework, which are explained in section \ref{sec:ThermoTransport}. Reactions with defined energy exchange ratios, which may lead to enthalpy disparities (e.g., intermediate species production), are balanced by redistributing energy across other modes when feasible \cite{pokharelSelfconsistentModelNumerical2023a}. This approach ensures a conservative energy balance for all modes, while also seamlessly integrating fast-gas heating processes.
Three-body reactions involving electrons as the third body contribute to electron heating. Vibrational excitation of ground-state molecular nitrogen through chemical reactions \cite{popov_associative_2009} is considered in reactions R14, R15, R36, R37, R185, and R186. Fast gas heating resulting from electron impact dissociation is explained by the formation of an intermediate excited species. Therefore, heating values for fast-gas heating in reactions R51, R191, and R192 are determined based on Popov's work \cite{popovFastGasHeating2011}. In addition to heat exchange in chemical reactive processes, other energy exchange terms $Q_{ET}$, $Q_{EV}$, $Q_{VE}$, and $Q_{VT}$ are presented in Eq. \ref{eq:Qs}, and they represent the heating rate due to elastic electron-molecule collisions, electron to vibrational excitation, vibrational relaxation to electrons, and vibrational relaxation to heavy particles, respectively.

\begin{align}
    \nu_{ei} = 3.636 \times 10^{-6} n_e T_e^{-3/2} & \ln (\Lambda); \notag \\
    \ln(\Lambda) &= 0.5 \ln \left( 1 + \frac{\lambda_D^2}{r_{\text{ref}}^2} \right)
    \label{eq:nu_ei}
\end{align}

Here, $\nu_{en}$ and $\nu_{ei}$ represent the collision frequencies of electrons with neutrals and ions, respectively. $\nu_{en}$ is calculated based on collision cross-sections using BOLSIG+, while the Coulomb collision frequency is approximated with Eq. \ref{eq:nu_ei} \cite{kennedy_plasma_2021}, where $\ln(\Lambda)$ is the Coulomb logarithm. For a vibrational level $l$, $U_l$ denotes the excitation energy, $K_l$ signifies the rate of vibrational excitation, and $K_l^{inv}$ stands for the rate of vibrational to electronic relaxation, determined through the detailed balance principle using BOLSIG+. The first eight vibrational levels of \Ntw are considered for the exchange processes $Q_{EV}$ and $Q_{VE}$. Furthermore, $n_v$ indicates the total number density of the corresponding species included for vibrational excitation, $n_{vl}$ denotes the number density of vibrationally excited species at level $l$, $\tau_{VT}$ represents the effective vibrational-translational relaxation time based on the Landau-Teller formalism, and $E_v$ denotes the total vibrational energy per unit volume, while $\theta_v$ refers to the characteristic vibrational temperature. The relaxation time undergoes significant changes for collisions with atoms. Thus, the effective relaxation time in Eqn. \ref{eq:Qs} is computed as $\tau_{VT} = (\sum_r X_r) \cdot \left(\sum_r X_r / \tau_r\right)^{-1} $, with the coefficients for determining individual relaxation times for various colliding partners obtained from Park \cite{parkReviewChemicalkineticProblems1993}.

To prevent the Coulomb logarithm from becoming negative in laser-induced low-temperature plasmas, which occurs when \( r_0 > \lambda_D \) in conventional formalism where \( r_0 = e^2 / (4 \pi \varepsilon_0 m_e v_{\text{th}}^2) \) is the impact parameter and \( \lambda_D = k_B T_e / (4 \pi e^2 n_e) \) is the Debye length, we use the formulation proposed by Gericke et al. \cite{gericke_dense_2002} based on hyperbolic trajectories (Eq. \ref{eq:nu_ei}). This defines \( r_{\text{ref}} = (\lambda^2 + r_0^2)^{1/2} \), where \( \lambda = h / \sqrt{2 \pi m_e k_B T_e} \) is the thermal de Broglie wavelength.

\subsubsection{ Model for Laser Coupling}

Electrons in an electric field absorb energy with subsequent collisions through inverse-Bremsstrahlung absorption. A semiclassical treatment from Zeldovich and Raizer \cite{zeldovichPhysicsShockWaves2002}, Eq \ref{eq:QaE}, is used here for the rate of absorption of energy per electron.

\begin{align}
    Q_{a,e} = \frac{e^2 I_L (\nu_{en} + \nu_{ei})}{m_e c \epsilon_0 \left( \omega_L^2 + (\nu_{en} + \nu_{ei})^2 \right)}
    \label{eq:QaE}
\end{align}

\begin{subequations}\label{eq:ionization_frequency}
\begin{align}
\nu_i \approx& \alpha^2 \beta \xi \nu_E; \quad 
\xi  \approx 2 \alpha \exp\left( - \frac{\alpha - 1}{\alpha} \left( \frac{6 \nu^*}{\nu_E} \right)^{1/2} \right) \\
\nu_{i,tot} &= Q_{a,e} \beta \sum_{k} \frac{\eta_k X_k}{\tilde{I}_{p,k}}
\end{align}
\end{subequations}

\begin{equation}
\label{eq:net_laserAbsorb}
Q_a = n_e \left( Q_{a,e} - Q_{a,e} \beta \sum_{k} \frac{\eta_k X_k}{\tilde{I}_{p,k}} I_{p,k} \right)~~~[Jm^{-3}s^{-1}]
\end{equation}

Starting from the kinetic equation for the electron energy distribution function, Raizer \cite{raizerGasDischargePhysics1997a} derives an approximate formulation for avalanche ionization frequency due to energy absorption through inverse Bremsstrahlung. For avalanche ionization, we consider multi-step ionization resulting from ionization of \Ntw, \Otw, and \NO. We define the energy-gaining frequency based on the time required for electrons to gain slightly more energy than ionization potential ($\approx$ 1 eV above, denoted $\tilde{I}_{p}$), which gives $\nu_E = Q_{a,e}/\tilde{I}_p$. Raizer \cite{raizerGasDischargePhysics1997a} provides two solutions for ionization frequency for cases, $\nu^* < \nu_E$ and $\nu^* > \nu_E$, where $\nu^*$ represents the threshold energy for excitation. When the collisional loss of energy during energy absorption is significant (i.e., $\nu^*> \nu_E$), the avalanche ionization frequency changes, as the loss impedes ionization (see Eq. \ref{eq:ionization_frequency}). $X_k$ is the mole fraction of species $k$ while $\alpha^2 = \tilde{I}_{p}/E^*$ (the fraction of ionization to the excitation thresholds). The parameter $\beta$ is an empirical parameter representing the probability of ionization when the electron gains energy $\tilde{I}_{p}$ from the electric field. Here, we take $\beta = 0.95$. At high intensities, we limit $\eta_k = 1$, thereby reproducing the limiting solution for $\nu^* < \nu_E$, i.e., $\nu_i \approx \beta \nu_E$. The parameters used for determining the avalanche ionization frequency include excitation threshold energies of $12.03~\text{eV}$ for $N_2$ and $9.40~\text{eV}$ for $O_2$, along with ionization thresholds of $16.58~\text{eV}$ for $N_2$ and $13.06~\text{eV}$ for $O_2$ (note 1 eV more than the potential). Additionally, the averaged excitation rate for both species is considered to be $10^{-9}~\text{cm}^3/\text{s}$ based on results from BOLSIG+. $\eta_{NO}$ was taken to be equal to $\eta_{O_2}$. The species production rate from avalanche ionization, calculated as $n_e \cdot \nu_i$, is incorporated as a source term in the corresponding species conservation equation. The net energy absorbed, $Q_a$, as defined in Eq. \ref{eq:net_laserAbsorb}, is included as a source term in the electron energy equation.

\begin{figure}
    \centering
    \includegraphics[width=0.95\linewidth]{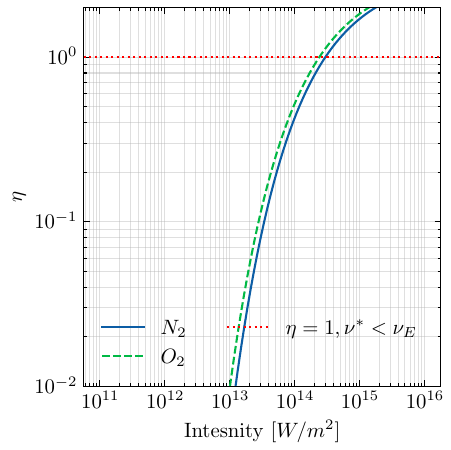}
    \caption{Avalanche ionization probability ($\eta$) as defined in Eq. \ref{eq:ionization_frequency}.}
    \label{fig:probeNs}
\end{figure}

Electron photodetachment induced by laser radiation contributes significantly to both electron generation and the loss of negative ions. Specifically, we consider photodetachment from \Otwm and neglect those from \Om (threshold energy $\approx$ 1.4 eV) \cite{Shneider2011TailoringPulse} since the heating laser considered in this study operates at 1064 nm ($E_{ev} \approx$ 1.16 eV). The cross-section for this process is described by $\sigma_{pd} = 10^{-4} \cdot E_{ev}(E_{ev}-0.15)^{1.5} \left( 0.370 \times 10^{-17} + 0.071 \times 10^{-17} (E_{ev}-0.15) \right)~[m^2]$ \cite{burchPhotodetachmentMathrmEnsuremath1958}. The corresponding source term in the species conservation equation has the form as shown in Eq. \ref{eq:pdRates}.

\begin{equation}\label{eq:pdRates}
\frac{d n_e}{dt}\bigg|_{pd} = -\frac{d n_{O_2^-}}{dt}\bigg|_{pd} = \sigma_{pd} \frac{I_L}{E_{J}} N_{O_2^-}
\end{equation}

The Keldysh parameter, or adiabaticity parameter, $\gamma_k$, is defined as the ratio of the laser frequency to the frequency of electron tunneling through the potential barrier, given by $ \gamma_k = {\omega \sqrt{2 m_e I_p}}/{e E} $, where $\omega$ represents the laser frequency, $I_p$ is the ionization potential, and $E$ denotes the electric field strength \cite{keldysh_ionization_1964, perelomov_ionization_1966}. Our study employs the Perelomov-Popov-Terent'ev (PPT) theory\cite{perelomov_ionization_1966} to determine the photo-ionization rate ($\Omega_{h\nu}$) in air under irradiation by a plane-polarized laser field. The intensity of the 1064 nm heating pulse is small for any noticeable photo-ionization. In the dual-pulse setup considered in this study, pre-ionization is generated using a fs laser with a wavelength of $\lambda = 800~\text{nm}$, a pulse width of $\Delta t_L = 100~\text{fs}$, and a beam waist radius of $\omega_0 = 150~\mu\text{m}$. The ionization yield is calculated based on the clamped laser intensity, determined by the balance between Kerr focusing ($\Delta n = \eta_2 I_{cl}$) and plasma defocusing \cite{kasparian_critical_2000}. The calculated clamped intensity is $I_{cl} \approx 5.469 \times 10^{17}~\text{W/m}^2$ for air and $I_{cl} \approx 6.44 \times 10^{17}~\text{W/m}^2$ for nitrogen at atmospheric pressure.

\subsection{Validation of the models}

\subsubsection{Plasma Kinetics}


\begin{figure*}
    \centering
    \subfloat[Comparison with experiments\label{fig:valAir}]{%
        \includegraphics[width=0.45\textwidth]{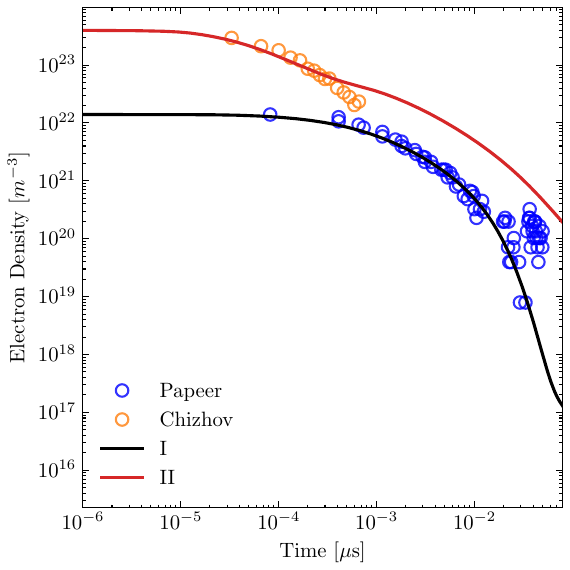}%
    }
    \hspace{0.075\textwidth}%
    \subfloat[Comparison with various models models\label{fig:compN2}]{%
        \includegraphics[width=0.45\textwidth]{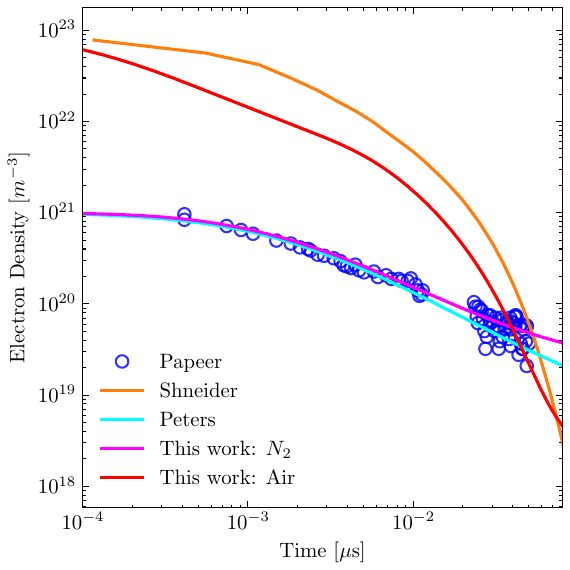}%
    }
    \caption{Temporal evolution of electron density decay in fs-laser induced 
             plasma filament at atmospheric pressure. Comparisons with 
             Papeer \cite{papeer_temporal_2013}, 
             Chizov \cite{chizhov_interferometry_2015}, and 
             Peters \cite{peters_considerations_2019,pokharelSelfconsistentModelNumerical2023a}.}
    \label{fig:valPlasma}
\end{figure*}

The plasma transport and hydrodynamics approach of the solver was validated in previous work \cite{pokharelSelfconsistentModelNumerical2023a}. In this study, the extended plasma kinetics model is independently validated for air and pure nitrogen mixtures, as shown in Fig. \ref{fig:valPlasma}. Note that in both cases, the same full reaction set is utilized. To validate the developed plasma kinetics model, initial conditions reflecting experimental data were employed. The intensity of the operating fs-pulse (100 fs, 800 nm) was varied to attain the desired total electron density at the end of the fs-pulse. This ensures a consistent relative population of ions, specifically \Ntwp and \Otwp for air after fs preionization.

The method described by Papeer \cite{papeer_temporal_2013} measures electron density by observing the reduction in microwave intensity as it propagates through the plasma, with the signal attenuation being proportional to the electron density. Chizov \cite{chizhov_interferometry_2015} employed optical interferometry to estimate electron density by measuring changes in the plasma's refractive index. Comparison of the experimental results with the simulation results from this work shows good agreement in electron density decay profiles in pure nitrogen and air.

In addition to the comparison of the plasma decay in nitrogen (\Ntw) with Papeer's experimental measurements \cite{papeer_temporal_2013}, we also compared our results with the plasma kinetics model developed by Peters \cite{peters_considerations_2019,pokharelSelfconsistentModelNumerical2023a} for pure nitrogen, which has been revised and extended in this study. The simulation results demonstrate that, while the overall decay of electrons remains largely unchanged, there is a noticeable slower plasma decay at longer times ($t > 10 ns$) captured by the revised model. This is attributed to the incorporation of associative ionization processes, which involve electronically excited atomic nitrogen in the current work, as explained by Popov\cite{popov_associative_2009}. It should be noted that such modification of the kinetics model allows us to  capture electron decay dynamics more accurately, consistent with the experimental results of Papeer \cite{papeer_temporal_2013} (see Fig. \ref{fig:compN2}).

\subsubsection{Laser coupling and breakdown}

\begin{figure}
    \centering
    \includegraphics[width=0.95\linewidth]{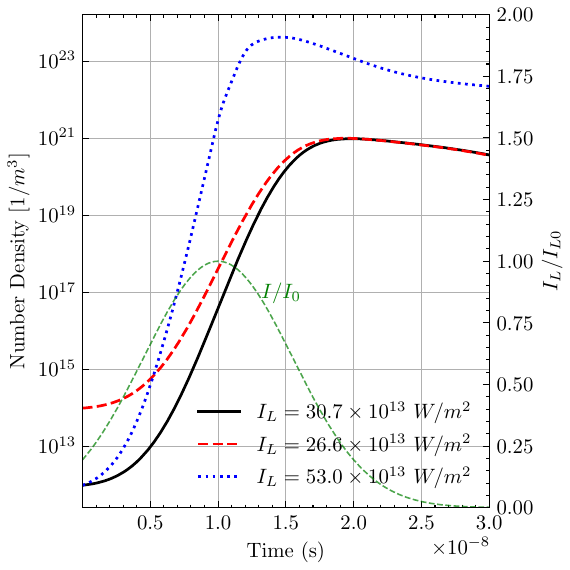}
    \caption{Temporal evolution of the electron density at various intensities of the ns-pulse, illustrating breakdown and decoupling thresholds in air.}
    \label{fig:breakdownComp}
\end{figure}

To validate the laser absorption part of the model, we refer to Yalin's experiments and modeling work on UV preionization for double-pulses \cite{yalinLaserPlasmaFormation2014}. The breakdown threshold criterion in the reference paper is based on the critical electron density of $10^{21}$ m\textsuperscript{-3}. Although quantitative experimental measurements of the electron density were not reported, the work provided the experimental threshold intensity of breakdown (observed through strong light emission and sound) along with theoretical modeling results. Our comparison focuses on a single ns-pulse with a pulse width of 13 ns, utilizing a 1064 nm pulse. While Yalin \cite{yalinLaserPlasmaFormation2014} employed a constant probability of avalanche ionization and a constant intensity model, we incorporate the temporal progression of the pulse intensity to ensure a smooth transition from low to high levels of avalanche ionization, by accounting for instantaneous changes in the probability of ionization based on the temporal intensity profile of the laser. We consider two initial electron densities of $10^{12}$ and $10^{14}$ m\textsuperscript{-3} and include the possibility of ionization from \NO. The simulation results for the breakdown threshold intensity are shown in Table \ref{tab:yalinComp} and Fig. \ref{fig:breakdownComp}. Our findings align well with the results reported by Yalin, showing minimal variation with the initial electron density.

\begin{table}
    \caption{Threshold intensity for a single ns-pulse (1064 nm) breakdown ($n_e = 10^{21}~\text{m}^{-3}$).}
    \begin{ruledtabular}
    \begin{tabular}{lcl}
        \textbf{$n_{e0}$} $[\text{m}^{-3}]$ & \textbf{$I_{n0}$} $[\text{GW}/\text{cm}^2]$ & \textbf{Remarks} \\
        \hline
        $10^{12}$ & 30.7 & This work \\
        $10^{14}$ & 26.60 & This work \\
        $10^{14}$ & 26.0 & Yalin \cite{yalinLaserPlasmaFormation2014} (Modeling) \\
        -- & 17.90 & Yalin \cite{yalinLaserPlasmaFormation2014} (Experiment) \\
    \end{tabular}
    \end{ruledtabular}
    \label{tab:yalinComp}
\end{table}

The previously defined breakdown criterion, based on a critical electron number density of $1 \times 10^{21}$ m\textsuperscript{-3}, provides limited insight into the threshold intensity for controlled laser energy deposition. Further controlled energy addition over plasma of $10^{21}$ m\textsuperscript{-3} is still possible using a heating pulse. To establish a threshold beyond which additional energy deposition from the second pulse leads to a loss of control over the energy deposition process, we introduce a decoupling threshold. The objective is to achieve controlled energy deposition, ensuring consistent and controlled plasma properties such as electron density, vibrational temperature, and gas temperature, without excessively heating the mixture ($T_g < 10000$ K). Thus, the aim is to determine the threshold at which the transition to uncontrolled absorption and breakdown occurs.

The decoupling threshold electron density is defined as the peak electron density at the plasma channel core, where the resultant refractive index gradient causes the laser beam to defocus, producing a deviation angle exceeding three degrees within a short distance ($\Delta x \ll L_p$). This short distance is taken as approximately 20 times the filament radius. The Eikonal approximation of the wave equation \cite{Born1999PrinciplesOptics,zhaoFastSweepingMethod2005} is employed to determine the required refractive index at the core, assuming a linear gradient. For a plasma filament radius of 50 $\mu$m and a wavelength $\lambda = 1064$ nm, the threshold electron density is approximately $4 \times 10^{23}~\text{m}^{-3}$, corresponding to the required intensity of $53 \times 10^{13}~\text{W/m}^2$ at the core, as shown in Fig.~\ref{fig:breakdownComp}. At these conditions the peak bulk gas temperature reaches $\approx$ 6000 K.

\section{Results and Discussion}

In this section, we provide a detailed analysis of the temporal evolution of laser-induced plasma, focusing on the decay of fs-laser-generated plasma in air at atmospheric pressure to identify the dominant pathways and key reactions. We further assess the impact of oxygen addition in N$_2$-O$_2$ plasmas and examine the spatio-temporal evolution of laser plasma across various initial electron densities and temperatures. Additionally, we investigate the concept of dual-pulse techniques, specifically the coupling of laser energy from a ns heating pulse to the fs-laser induced filament. Particular attention is given to the temporal shaping of the ns heating pulse and its influence on plasma dynamics, as well as the efficiency of dual-pulse laser energy deposition.

\subsection{Problem setup and initialization}

\begin{figure}
    \centering
    \includegraphics[width=0.85\linewidth]{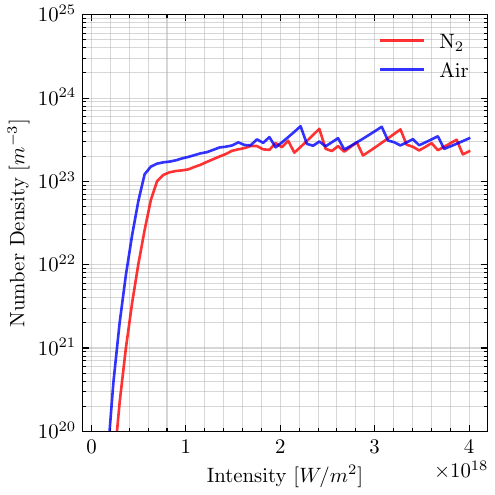}
    \caption{Number density of electrons generated in the fs-laser fialment in nitrogen and air. Tighter focusing results in saturation at higher electron density.}
    \label{fig:fsYield}
\end{figure}

Accurate determination of properties of the fs-laser filament requires modeling of the fs laser pulse propagation, which incorporates a nonlinear Kerr focusing effect, mutiphoton and tunneling ionization, electronic excitation, wave dispersion, multi-filamentation and other optical phenomena. Although a detailed analysis of the fs laser pulse propagation is beyond the scope of this work, previous studies \cite{thebergePlasmaDensityFemtosecond2006, bakTwodimensionalHighResolution2024, aleksandrov_decay_2016}, have established some properties of fs-laser-induced filaments, particularly electron density, while reliable measurements of electron temperature and other species at the end of the pulse remain limited. The electron density in such filaments is primarily affected by the focusing setup and exhibits a saturation effect on the electron yield with the increase of the fs-laser energy \cite{Theberge2006PlasmaFocusing}. Based on the saturation effect, we use the measured filament diameter ($\approx 100 ~\mu m$) observed in experiments to determine the decoupling threshold electron density. In the moderate focusing regime considered, the local intensity within the filament corresponds to the base clamping intensity (for self-propagating) augmented by an additional intensity, such that the final yield of electrons density at high energy saturates near the decoupling threshold. The clamped intensity and the corresponding electron density yield, based on the PPT theory, are detailed in \cite{pokharelSelfconsistentModelNumerical2023a}. Figure~\ref{fig:fsYield} shows the electron density at the end of the fs-pulse for both air and nitrogen at 1 atm. Following ionization, free electrons absorb additional photons, and a simplified energy balance analysis is used to calculate the electron temperature as $T_e = (2/3) (h \nu_l (N_{ph} + N_{ph,+}) - I_p)$. Here, $N_{ph}$ denotes the photons required for ionization, and $N_{ph,+}$ represents the additional absorbed photons. Assuming excess absorption of 2 and 4 photons in air-plasma, the resulting electron temperatures are 2.14~eV and 4.20~eV, respectively. The vibrational temperature is modeled solely for molecular nitrogen, and excludes other species.

\subsection{Air Plasma Decay \label{sec:airPlasmaDecay}}


The plasma decay dynamics in air are analyzed starting from the parameters of the fs laser filament at the end of the fs pulse. At the clamping laser intensity of $\approx 6 \times 10^{17}$ W/m$^2$, an electron yield of $ \approx  10^{23}$ m$^{-3}$ is obtained, and an electron temperature ($T_e$) is set at $3$ eV. After the end of the fs laser pulse, processes such as recombination, gas heating, and energy exchanges between various energy modes occur. Once sufficient thermalization is achieved, hydrodynamic expansion of the formed plasma channel occurs, which eventually cools over time due to diffusion processes.

\begin{figure}[]
    \centering
    \includegraphics[width=0.95\linewidth]{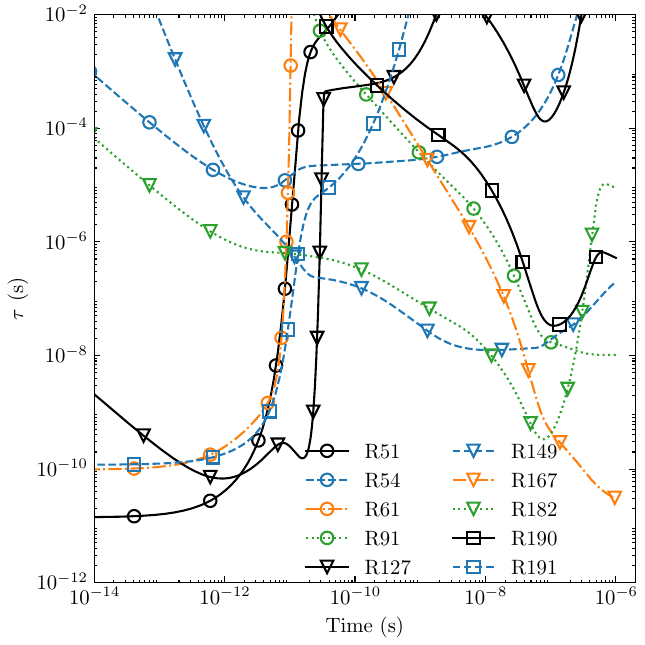}
    \caption{Temporal scales of main reactions in the fs-laser induced plasma in air at atmospheric pressure.}
    \label{fig:tauImp}
\end{figure}

\begin{figure}[]
    \centering
    \includegraphics[width=0.95\linewidth]{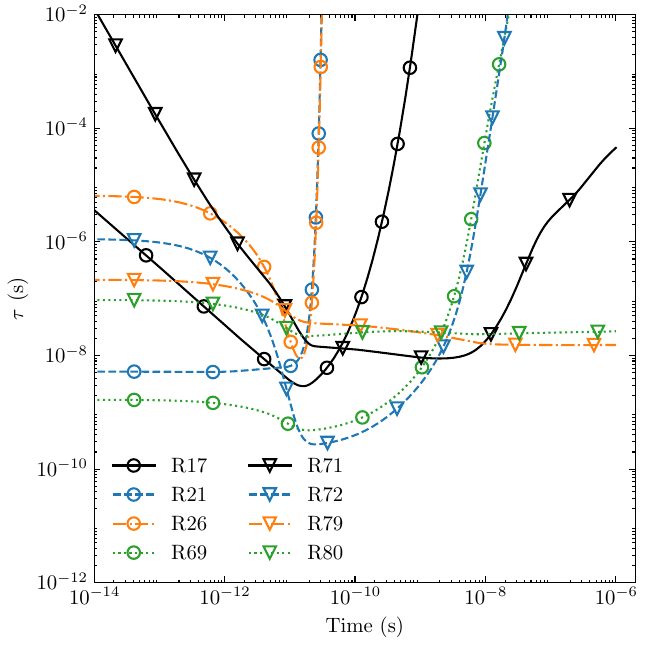}
    \caption{Temporal scales of main reactions involving the loss of electrons in the fs-laser-induced plasma in air at atmospheric pressure.}
    \label{fig:tauEloss}
\end{figure}


To characterize the overall kinetic behavior of the plasma, the temporal scales of various collisional processes are evaluated. The characteristic timescale ($\tau$) is determined using the chemical kinetic rate expression $\tau = (\omega_k/n_e)^{-1}$, where $\omega_k$ denotes the reaction rate of the $k$th chemical process. The results of this analysis are presented in Fig.~\ref{fig:tauImp} for the main reactions in air plasma and in Fig.~\ref{fig:tauEloss} for electron energy loss reactions, with detailed descriptions of the specific reactions provided in Table~\ref{tab:ImpRxns}. The key kinetic processes governing FLI plasma decay in air are identified as follows:

\begin{itemize}
    \item Elastic collisions between electrons and heavy particles ($Q_{ET}$) occur rapidly with $\tau \approx 10^{-12} \, \text{s}$, as do vibrational excitations ($Q_{EV}$) with $\tau \approx 10^{-11} \, \text{s}$.
    \item Initially, electron impact dissociation of neutrals is significant (e.g. R51, R191), with a high rate at early times, $\tau \approx 10^{-11} - 10^{-10} \, \text{s}$, leading to a buildup of atomic nitrogen and oxygen.
    \item Cluster ion formation of \Nforp peaks around 10 ps, coinciding with the peak formation of \NOp (R127).
    \item Dissociative recombination of \Ntwp and \Otwp along with electronic excitation of molecular nitrogen occur within $\tau \approx 10^{-10}-10^{-9} \, \text{s}$. Note that for pure nitrogen the recombination with \Nforp is the main channel for decay at ns timescales thereby producing high population of \NtwB but this is weak in air-plasma because of less population of \Nforp.
    \item Attachment operates on a timescale of about $\tau = 10^{-8} \, \text{s}$, remaining relatively constant throughout the decay process. Electron recombination with \NOp also has a $\tau = 10^{-8} \, \text{s}$ timescale, but only lasts for tens of nanoseconds. 
    \item $\mathrm{NO}$ formation begins slowly through ion-ion neutralization of $\mathrm{NO}^+$ and $\mathrm{O}_2^-$ (R149) until 10 ns. After 10 ns, with sufficient accumulation of electronically excited $\mathrm{N}(D)$, rapid $\mathrm{NO}$ production occurs via $\mathrm{N}(D)$ recombination (R182). $\mathrm{NO}$ is primarily consumed through dissociation (R167) after 100 ns.
    \item Detachment becomes prominent after significant attachment and operates on a $\tau = 10^{-8} \, \text{s}$ timescale but emerges around 100 ns. At around 100 ns, the relaxation of electronically excited atoms and associative ionization occur, resulting in a decrease in density of the excited atoms.
    \item Various atomic recombination processes become significant after several hundreds of nanoseconds.

\end{itemize}

Thus, plasma decay dynamics can be divided into early ($t < 10^{-11} \, \text{s}$), mid ($10^{-11} \, \text{s} < t < 10^{-9} \, \text{s}$), and late ($t > 10^{-9} \, \text{s}$) stages, based on the timescales of various processes, as illustrated in Fig. \ref{fig:tauEloss}. In the early stage, two-body dissociative recombination predominates (R69, R21, R22), while dissociative recombination from clusters and three-body recombination with \Otw (R72, R75) dominate the mid-stage. Two-body dissociative recombination of \Otwp (R69) significantly affects both early and mid-stages. The late-stages begin with a prominent recombination with \NOp (R71), followed by attachment reactions (R79, R80), which drive  the remainder of the late stage. In air plasma, electron production beyond 10 ns is primarily controlled by detachment in collisions with oxygen (R91) and associative ionization (R190), with characteristic timescales of approximately 10 ns, which slows down the plasma decay. The main reactions depicted in the figures   \ref{fig:tauImp} and \ref{fig:tauEloss} are summarized in table \ref{tab:ImpRxns}, and the complete set of reactions is presented in the appendix.

\begin{table}[htbp]
    \caption{Main reactions in air-plasma as presented in Figures \ref{fig:tauImp} and \ref{fig:tauEloss}.}
    \begin{ruledtabular}
    \begin{tabular}{lp{3.25cm}lp{3.4cm}}
        \textbf{ID} & \textbf{Reaction} & \textbf{ID} & \textbf{Reaction} \\
        R17 & $\mathrm{N}_4^+ + \mathrm{e}^- \rightarrow \mathrm{N}_2 + \mathrm{N}_2(\mathrm{B})$ & R21 & $\mathrm{N}_2^+ + \mathrm{e}^- \rightarrow \mathrm{N} + \mathrm{N}(\mathrm{D})$ \\
        R26 & $\mathrm{N}_2^+ + 2 \mathrm{e}^- \rightarrow \mathrm{N}_2 + \mathrm{e}^-$ & R51 & $\mathrm{N}_2 + \mathrm{e}^- \rightarrow \mathrm{N} + \mathrm{N}(\mathrm{D}) + \mathrm{e}^-$ \\
        R54\footnote{$\mathrm{v \in [0, 4]}$. See the complete reaction mechanism for details.} & $\mathrm{N}_2(\mathrm{A},v) + \mathrm{e}^- \rightarrow \mathrm{N}_2 + \mathrm{e}^-$ & R61 & $\mathrm{N}_2 + \mathrm{e}^- \rightarrow \mathrm{N}_2(\mathrm{C}) + \mathrm{e}^-$ \\
        R69 & $\mathrm{O}_2^+ + \mathrm{e}^- \rightarrow 2 \mathrm{O}$ & R71 & $\mathrm{NO}^+ + \mathrm{e}^- \rightarrow \mathrm{N} + \mathrm{O}$ \\
        R72 & $\mathrm{O}_2^+ + 2 \mathrm{e}^- \rightarrow 2 \mathrm{O} + \mathrm{e}^-$ & R79 & $2 \mathrm{O}_2 + \mathrm{e}^- \rightarrow \mathrm{O}_2 + \mathrm{O}_2^-$ \\
        R80 & $\mathrm{N}_2 + \mathrm{O}_2 + \mathrm{e}^- \rightarrow \mathrm{N}_2 + \mathrm{O}_2^-$ & R91 & $\mathrm{O} + \mathrm{O}_2^- \rightarrow \mathrm{O}_3 + \mathrm{e}^-$ \\
        R127 & $\mathrm{N}_2^+ + \mathrm{O} \rightarrow \mathrm{N} + \mathrm{NO}^+$ & R149 & $\mathrm{N}_2 + \mathrm{NO}^+ + \mathrm{O}_2^- \rightarrow \mathrm{N}_2 + \mathrm{NO} + \mathrm{O}_2$ \\
        R167 & $\mathrm{N} + \mathrm{NO} \rightarrow \mathrm{N}_2 + \mathrm{O}$ & R182 & $\mathrm{N}(\mathrm{D}) + \mathrm{O}_2 \rightarrow \mathrm{NO} + \mathrm{O}(\mathrm{D})$ \\
        R190 & $\mathrm{N}(\mathrm{P}) + \mathrm{O} \rightarrow \mathrm{NO}^+ + \mathrm{e}^-$ & R191 & $\mathrm{O}_2 + \mathrm{e}^- \rightarrow 2 \mathrm{O} + \mathrm{e}^-$ \\
    \end{tabular}
    \end{ruledtabular}
    \label{tab:ImpRxns}
\end{table}

\begin{figure*}[]
    \centering
    \includegraphics[width=0.65\linewidth]{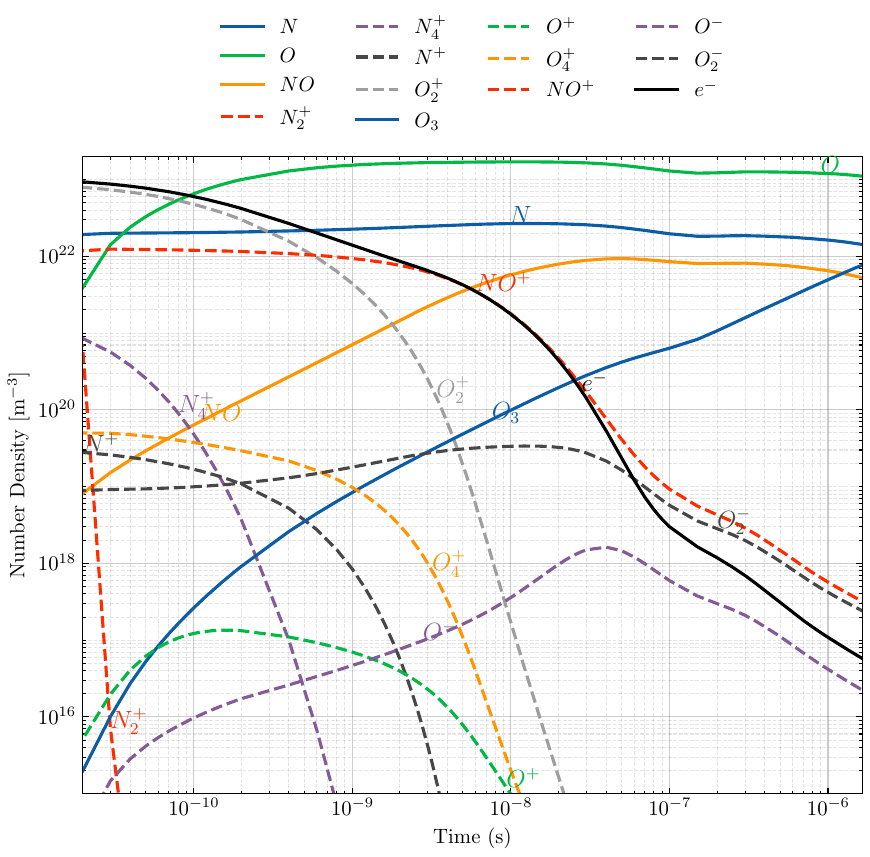}
    \caption{Temporal evolution of number densities of various species, including ions, at the center of the fs-laser induced plasma in air at atmospheric presure.}
    \label{fig:probeDecay0}
\end{figure*}

\begin{figure*}[]
    \centering
    \includegraphics[width=0.65\linewidth]{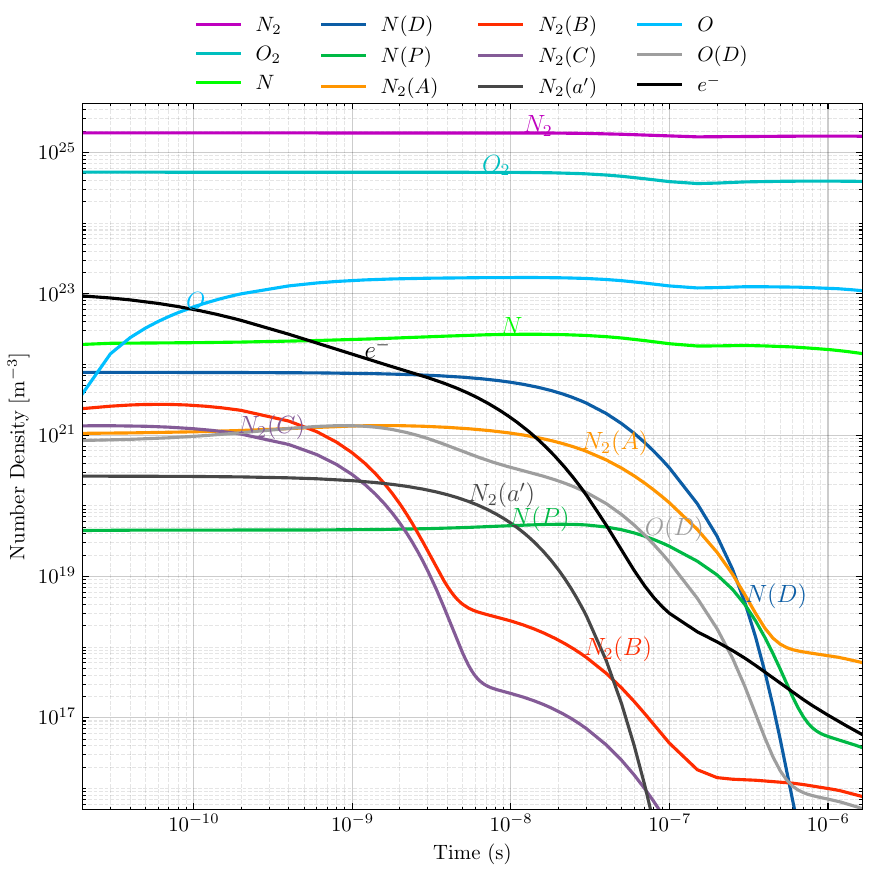}
        \caption{Temporal evolution of number densities of various species, including electronically excited states, at the center of the fs-laser induced plasma in air at atmospheric presure.}
    \label{fig:probeDecay1}
\end{figure*}

\begin{figure}[]
    \centering
    \includegraphics[width=0.95\linewidth]{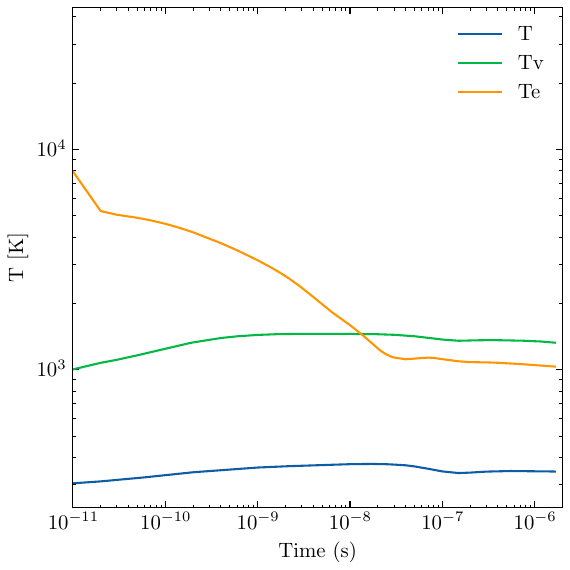}
    \caption{Temporal evolution of the bulk gas temperature $T$, vibrational temperature $T_v$, and electron temperature $T_e$ at the core of the fs-laser induced plasma channel in air.}
    \label{fig:probeTair}
\end{figure}

The temporal decay of various species (neutrals, electrons, ions, and electronically excited species) in the fs-laser induced filament in air at atmospheric pressure, taken at the core of the plasma channel, is illustrated in figures \ref{fig:probeDecay0} and \ref{fig:probeDecay1}. The subsequent temporal evolution of gas, vibrational, and electron temperatures is depicted in Fig.~\ref{fig:probeTair}. Observations indicate a higher concentration of atomic oxygen compared to atomic nitrogen during the early stages of plasma decay, primarily due to dissociative recombination and electron impact dissociation. Both processes result in the rapid production of atoms within nanoseconds for oxygen and even faster for nitrogen.

The maximum density of \(\text{NO}\) is reached at approximately 50 ns. Electronically excited states of nitrogen, particularly \(\text{N}_2(A)\) and \(\text{N}_2(B)\), show a rapid decrease within the first few nanoseconds, reaching low concentrations of about \(1.0 \times 10^{16} \, \text{m}^{-3}\) around 100 ns. For pure nitrogen, the decay timescale of the ground electronic state of atomic nitrogen reported in literature  \cite{nakagawaTimeEvolutionAtomic2022} is approximately 100 \(\mu\text{s}\). In air plasma, the decay occurs on the \(\mu\text{s}\) timescale due to faster quenching in the presence of oxygen. All electronically excited atomic states exhibit a sharp decay around 100 ns. A slight drop in the bulk density during shock detachment caused by the hydrodynamic expansion of the plasma channel is observed around 100 ns on the decay profiles of atomic and molecular nitrogen and oxygen.

\subsubsection{Analysis of energy exchange channels \label{sec:AnalysisEnergyExchange}}

With an initial electron temperature of 3 eV in the fs-laser filament, electron impact processes, including dissociation (R51) and electronic excitation (R61, R191), dominate the initial electron energy loss within the first 10 ps, while vibrational excitation ($Q_{EV}$) plays a minor role. After 10 ps, $Q_{EV}$ becomes the primary electron energy loss mechanism and partially restored by three-body dissociative recombination (R72), superelastic pooling from electronically excited \(\text{N}_2\) (R62), and energy transfer from vibrationally excited \(\text{N}_2\) ($Q_{VE}$). 

The electron temperature equilibrates with the vibrational temperature over time. Gas heating from electron impact dissociation and dissociative recombination peaks at 50 ns, followed by cooling due to the hydrodynamic expansion of the plasma channel. The development and radial propagation of the shock wave are seen in density profiles at various times. For the same initial electron density, the vibrational temperature of \(\text{N}_2\) is higher in air than in nitrogen, as the presence of atomic oxygen enhances the recombination of atomic nitrogen into vibrationally excited \(\text{N}_2\). The lower ionization potential of air further reduces energy absorption from the fs-laser pulse, leading to a lower peak gas temperature compared to nitrogen.

\begin{figure}[]
    \centering
    \includegraphics[width=0.95\linewidth]{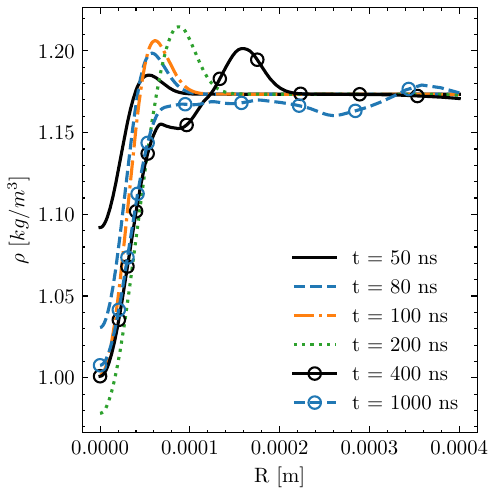}
    \caption{Radial distribution of the bulk gas density at various times during the decay of the fs-laser induced filament in air.}
    \label{fig:probeRhoAir}
\end{figure}

\subsection{Spatial profiles and refractive index dynamics}

\begin{figure*}
    \centering
    \subfloat[Radial distributions for \ele.\label{fig:radialEleAir}]{%
        \includegraphics[width=0.425\textwidth]{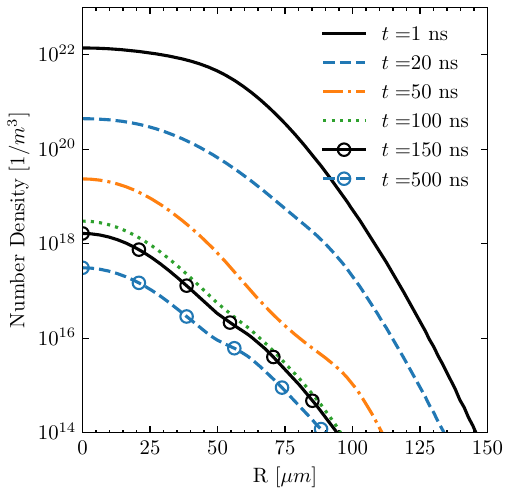}%
    }
    \hspace{0.075\textwidth}%
    \subfloat[Radial distributions for \Otwm.\label{fig:radialO2mAir}]{%
        \includegraphics[width=0.425\textwidth]{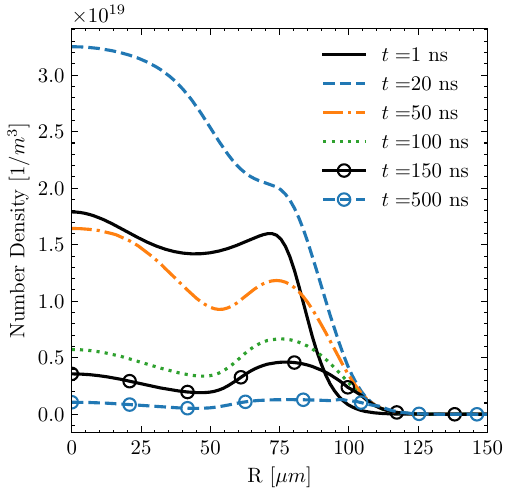}%
    }
    \caption{Radial distribution of number density of electrons and \Otwm 
             at various times for the fs-laser induced filament in air.}
    \label{fig:radialAir0}
\end{figure*}

\begin{figure}[]
    \centering
    \includegraphics[width=0.9\linewidth]{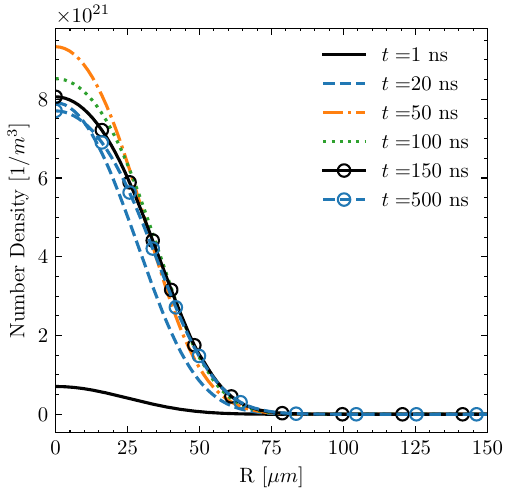}
    \caption{Radial distribution of the number density of \NO at various times for the  fs-laser induced filament in air.}
    \label{fig:radialNOAir}
\end{figure}

\begin{figure}[]
    \centering
    \includegraphics[width=0.9\linewidth]{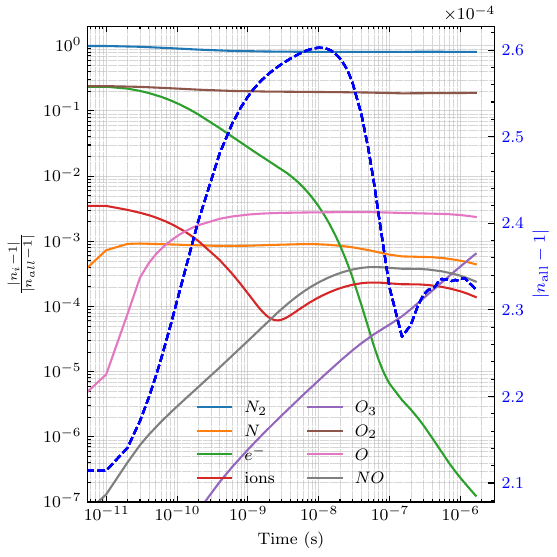}
    \caption{Refractive index contributions during the fs-laser plasma decay in air at atmospheric pressure for the wavelength of 1064 nm.}
    \label{fig:refraCompsAir}
\end{figure}

For multiple pulse applications and wave-guiding, the spatial distribution of species such as electrons, \(\text{O}_2^-\), and \(\text{NO}\) is critical for controlled coupling via spatially shaped pulses. Dynamics of the refractive index in the formed plasma filament is also critical for waveguiding and optical diagnostic applications. Thus, in this part, we concentrate on the radial distribution of species and corresponding changes in the temporal dynamics of the refractive index.
As shown in the temperature, Fig. \ref{fig:probeTair}, and bulk gas density profiles, Fig. \ref{fig:probeRhoAir}, thermalization induces a weak shock wave propagating radially outward,  The radial distribution of electrons and \(\text{O}_2^-\) in Fig. \ref{fig:radialAir0} and \(\text{NO}\) in Fig. \ref{fig:radialNOAir} shows different profiles. Electrons maintain a Gaussian-like central profile, expanding at the edges, while \(\text{NO}\) follows a near-Gaussian distribution. In contrast, \(\text{O}_2^-\) accumulates at the plasma edge. After 100 ns, the peak for \(\text{O}_2^-\) shifts from the center towards the radial edge. This local peak in \(\text{O}_2^-\) concentration at the radial interface results from the negative ion compression by ambipolar electric fields, as explained in \cite{pokharelSelfconsistentModelNumerical2023a}.

All these processes result in dynamical changes in the refractive index during the plasma decay in the fs-laser filament. Figure \ref{fig:refraCompsAir} shows the contribution of various species and processes in air refractivity. Initially, the refractive index is dominated by electron contributions, but a gradual shift to the hydrodynamic-dominant contribution is observed. A dip in the refractive index after the fs pulse is caused by the high electron density, after a rapid electron decay the refractive index returns to near-ambient levels at 10 ns. A reappearance of the dip occurs around 100 ns due to the plasma channel expansion. The contribution from dissociated oxygen exceeds that of nitrogen, highlighting the importance of considering dissociation in electron density measurements, particularly using interferometry \cite{pokharelRefractiveIndexModification2025}. The gradual increase in \(\text{O}_2^-\) contribution is attributed to its density growth and its threefold higher polarizability compared to molecular oxygen. Additionally, the contributions from ions and \(\text{NO}\) exceed those from electrons after about 30 ns.


\begin{figure*}
    \centering
    \subfloat[Degree of dissociation\label{fig:plasma_change_alpha}]{%
        \includegraphics[width=0.33\textwidth]{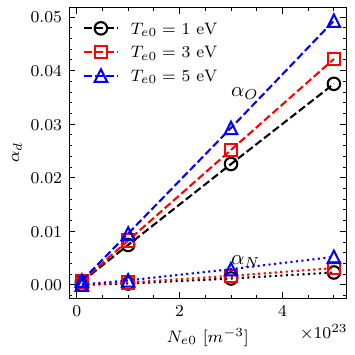}%
    }
    \hfill
    \subfloat[Peak vibrational temperature\label{fig:plasma_change_Tv}]{%
        \includegraphics[width=0.33\textwidth]{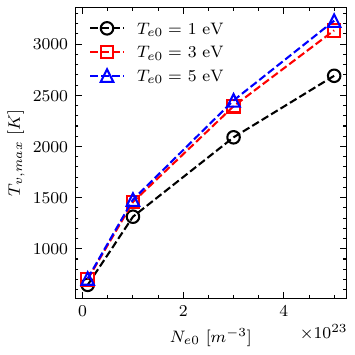}%
    }
    \hfill
    \subfloat[$T_{g,max}$ and density dip ratio\label{fig:plasma_change_thermal}]{%
        \includegraphics[width=0.33\textwidth]{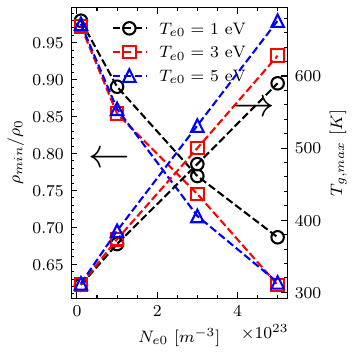}%
    }
    \caption{Characteristic properties of the fs laser filament in atmospheric 
             pressure air for various initial electron densities and electron 
             temperatures. }
    \label{fig:plasma_change_initials}
\end{figure*}

\subsection{Plasma characteristics at various initial conditions}

The characteristics of the fs laser filament, including its initial electron density and temperature, are primarily dictated by the laser focusing setup \cite{thebergePlasmaDensityFemtosecond2006}. In this analysis, electron density is considered within the range of \( 10^{22} \) to \( 5 \times 10^{23} ~ \text{m}^{-3} \) and electron temperature between \( 1 \) and \( 5 \) eV, with a focus on plasma decay properties and the influence of molecular oxygen (\(\text{O}_2\)). Plasma decay properties, including dissociation fraction, vibrational excitation, bulk density dip, and gas temperature, are examined and illustrated in Fig.~\ref{fig:plasma_change_initials}. The analysis reveals that the dissociation fraction of oxygen exceeds that of nitrogen significantly. The increase of the  electron density and temperature both enhance the rate of electron impact dissociation (see R51, R191). Moreover, the reaction's progression rate exhibits a linear dependence on the electron density. Elevated electron temperatures decrease rapidly due to energy loss from various channels, as previously discussed in section \ref{sec:AnalysisEnergyExchange}. For \( T_e > 1 \) eV, vibrational excitation rates plateau, and vibrational temperature saturates, similar to observations in nitrogen \cite{pokharelRefractiveIndexModification2025}, becoming independent of the initial electron temperature while varying with the initial electron density. A linear dependence of the bulk gas temperature on the initial electron density is observed with a stronger density dip due to higher thermalization, which is similar to the behavior observed during the decay of nitrogen plasma \cite{pokharelRefractiveIndexModification2025}.For identical plasma conditions, i.e., \( n_e \) and \( T_e \), the gas temperature in nitrogen plasma is higher than in air due to the greater energy required for nitrogen ionization. Gas temperature measurements in air-plasma filaments based on the N\textsubscript{2} second positive system reported temperatures in the range of 400–550 K \cite{edwards_simultaneous_2015,blanchardCharacterizationFemtosecondLaser2024}. In a filament with a tighter focusing setup, Blanchard \etal \cite{blanchardCharacterizationFemtosecondPlasma2025} measured a vibrational temperature of approximately 2700 K, which corresponds to an electron density of \(4.0 \times 10^{23} \, \text{m}^{-3}\) and \(T_e\) of 3 eV in Fig.~\ref{fig:plasma_change_initials}. Our recent Thomson scattering measurements \cite{urdanetaImplementationLaserThomson2024} showed that for an initial electron density of \(\approx 5.0 \times 10^{23} \, \text{m}^{-3}\), the gas temperature was about 700–800 K. A similar temperature of 700 K for \(T_e = 5\) eV is observed in our results for \(n_e = 5.0 \times 10^{23} \, \text{m}^{-3}\).

There is a possibility of vibrational excitation of molecular oxygen, with a much shorter VT relaxation time, about 100 ns, compared to nitrogen, which could explain the anomalous rise in the gas temperature around 100 ns\cite{urdanetaTemporallyResolvedProperties2025}. But an additional equation for the vibrational excitation of O\(_2\) has not been included in the model because of the absence any experimental measurements of the vibrational population of O\(_2\). However, the conclusions presented remain valid, as the temperature increase is significant only for cases with the high initial electron density. Even in these cases, while the temperature increase may be present due to the vibrational relaxation of O\(_2\), it does not exceed the peak thermalization temperature.

\begin{figure}[]
    \centering
    \includegraphics[width=0.9\linewidth]{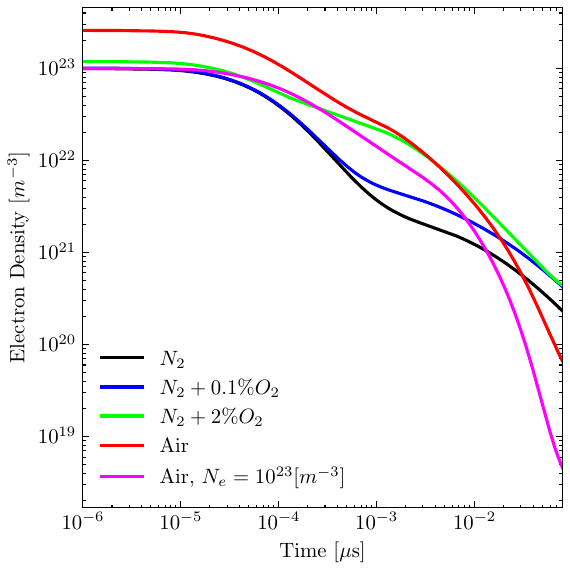}
    \caption{Electron density dependence on time in the fs-laser induced plasma in a mixture of \NtOt for various concentrations of \Otw at atmospheric pressure.}
    \label{fig:O2Changes}
\end{figure}

Now we assess the influence of oxygen on the plasma decay in the fs laser filament. Here, a laser pulse intensity of approximately \(7 \times 10^{17} ~ W/m^2\) is maintained, resulting in an electron density of \(10^{23}\) m$^{-3}$ at the end of the fs pulse for nitrogen (\(\text{N}_2\)), with oxygen concentrations adjusted to 0.1\%, 2\%, and 22\%. A comparison is also made with the plasma decay in air, ensuring a final electron density of \(10^{23}\) m$^{-3}$. Under these conditions, varying \(\text{O}_2\) concentration only slightly affects the electron density yield from the fs-laser. The temporal decay of electrons is shown in Fig.\ref{fig:O2Changes}. Nitrogen exhibits faster decay compared to air within the first few ns, given the same initial electron density. After 10 ns, electron attachment processes dominate in air, leading to a sharper plasma decay compared to that in nitrogen.

A sustained higher plasma density is observed for an extended time when a small percentage of \(\text{O}_2\) (around 2\%) is added. The main reason for this extension of the plasma life is that a significant portion of added \(\text{O}_2\) becomes ionized, resulting in a strong production of atomic oxygen from dissociative recombination. Since the molecular oxygen concentration is too small, attachment is not significant, because the main channel for attachment is a three-body process, \(2 \ \text{O}_2 + e^- \rightarrow \text{O}_2 + \text{O}_2^-\) with quadratic dependence on \Otw. As a result, attachment does not significantly affect the plasma decay in current conditions. At the same time, the presence of atomic oxygen enables associative ionization, such as \(\text{N}_2^+ + \text{O}_2 \rightarrow \text{NO} + e^-\), which slows down the plasma decay. As electron attachment processes intensify with the increase of \(\text{O}_2\), they eventually overcome the effect of associative ionization, resulting in the faster decay rates in air. Earlier experimental work \cite{bakCharacteristicsPlasmaDecay2024a} using microwave scattering to capture the dynamics of electron density has shown a similar trend of slowed decay with the addition of a small percentage of \Otw, highlighting the importance of associative ionization in air-plasma kinetics.

\subsection{Laser energy coupling to the fs laser filament}

Filaments induced by fs-lasers are repeatable and characterized by slender structures that can extend beyond the Rayleigh length. Controlled coupling of an ns heating pulse with these filaments offers potential for novel applications. The relevant findings are presented here. This section explores the plasma characteristics from both single laser pulse and dual laser pulse setups and identifies the main kinetic processes that characterize dual-pulse laser plasmas. The revival of the laser plasma using the dual-pulse approach, its dependence on the time delay between the two pulses, and the temporal shaping of the ns pulse intensity for optimal pairing of fs and ns pulses are also explored.

\subsubsection{Plasma generation with single pulse and dual Pulse}


\begin{figure*}
    \centering
    \subfloat[Single pulse (ns)\label{fig:SPcouple}]{%
        \includegraphics[width=0.46\textwidth]{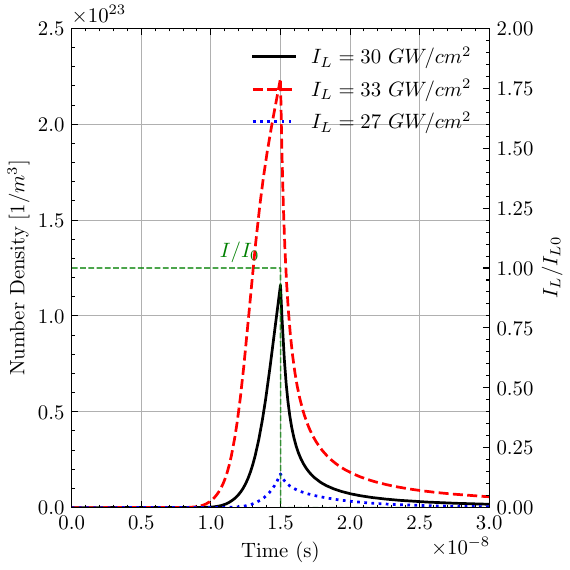}%
    }
    \hspace{0.05\textwidth}%
    \subfloat[Double pulse (fs-ns)\label{fig:DPcouple}]{%
        \includegraphics[width=0.46\textwidth]{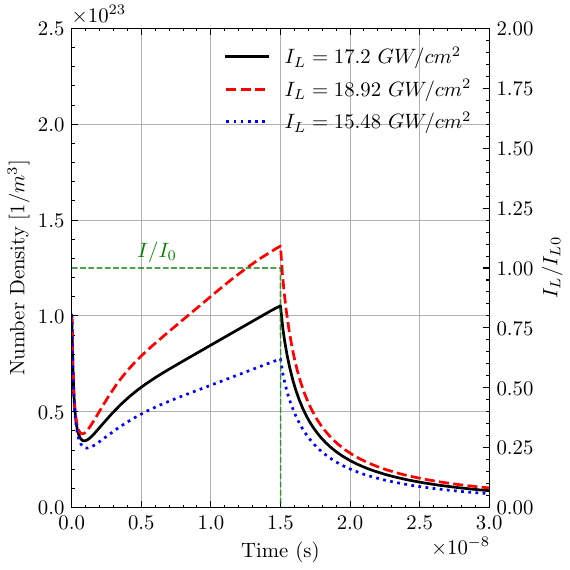}%
    }
    \caption{Laser-induced plasma formation from single ns-pulse and double 
             fs-ns pulse configurations.}
    \label{fig:SPvsDP}
\end{figure*}


Laser-induced plasma generates thermal, chemical, and hydrodynamic effects, important for various applications. The extent of thermalization depends on the ionization degree, with electron densities exceeding $10^{21}~m^{-3}$ producing significant effects. Here we focus on achieving an electron density of $10^{23}~m^{-3}$ and investigating the selectivity of the plasma properties and avoiding breakdown. The plasma is generated using both a single ns 1064 nm pulse (SP) and a dual fs-ns pulse (DP; 800 nm, 1064 nm). For simplification, the ns pulse is assumed to have a constant intensity and a 15 ns pulse duration. The fs pulse generates the plasma with $n_e = 10^{23}~\text{m}^{-3}$, $T_g = T_v = 300~\text{K}$, and $T_e = 3~\text{eV}$. This fs-plasma channel decays rapidly and requires revival to $n_e = 10^{23}~\text{m}^{-3}$ with the second heating pulse. In single ns-pulse operation, the electron density is achieved assuming  an initial electron density of $10^{12}~\text{m}^{-3}$ (see Fig. \ref{fig:breakdownComp} for cases with different initial electron densities).

The electron density time history for SP and DP is shown in figures \ref{fig:SPcouple} and \ref{fig:DPcouple}, respectively. The simulation results, shown in Fig. \ref{fig:SPvsDP}, indicate that the laser intensity required to achieve an electron density of $10^{23}~\text{m}^{-3}$ is 30 GW/cm\textsuperscript{2} for a single ns-pulse (SP) and 17.2 GW/cm\textsuperscript{2} for dual fs-ns pulses (DP). Additionally, the results in Fig. \ref{fig:SPvsDP} include a 10\% intensity variation for both configurations shown by the dashed and dotted lines. The DP configuration offers improved controllability over the SP setup, with smaller variations in electron density caused by fluctuations in operating conditions (modelled as a 10\% intensity variation). Temporal dynamics of the bulk gas and vibrational temperature (not shown) also lead to the same conclusion, with DP being less sensitive to variations in operating conditions. This results in better repeatability of plasma generation and more effective prevention of uncontrolled breakdown. The DP configuration also improves electron decay dynamics, leading to a slower decay rate after the second heating pulse, as seen in Fig. \ref{fig:DPcouple}.


\subsubsection{Main processes of dual-pulse laser energy coupling}

Fig. \ref{fig:DPcouple} shows a reduction in the plasma decay rate after revival for the DP configuration compared to the SP configuration, despite the same peak electron density. The key question is what kinetic processes are responsible for the slower plasma decay and the laser energy coupling by the ns laser pulse to the fs laser plasma. To answer this question, we evaluate the degree of the enhancement of the kinetic rates caused by the ns heating pulse utilizing the analysis of normalized rates of the electron production. These rates are defined by Eqn. \ref{eq:normalRates}.  It's worth noting that in this context, the SP refers to the plasma generation with a fs-laser pulse.

\begin{equation}
\text{[Normalized Rate]}_k = \left\| \frac{\left( \frac{dn_e}{dt}\big|_{DP} - \frac{dn_e}{dt}\big|_{SP} \right)_k}{\frac{dn_e}{dt} \big|_{SP}} \right\|
\label{eq:normalRates}
\end{equation}

\begin{figure}[]
    \centering
    \includegraphics[width=0.9\linewidth]{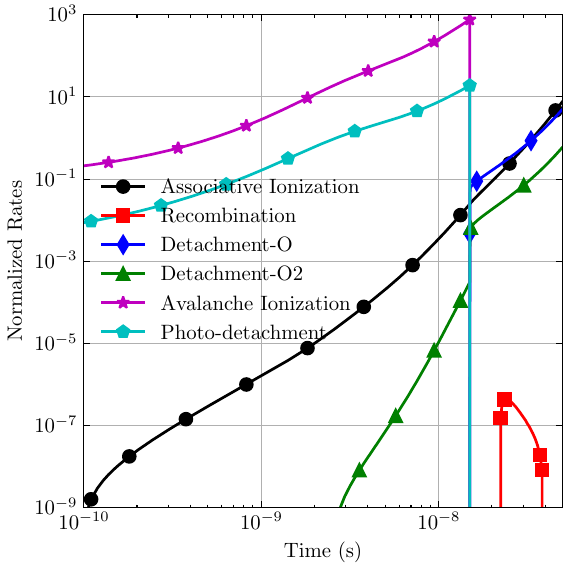}
    \caption{Normalized rates (Eq. \ref{eq:normalRates}) for various electron source terms for a dual-pulse laser plasma at the laser intensity of 17.6 GW/cm\textsuperscript{2}.}
    \label{fig:resCoupleDP}
\end{figure}

For a single fs-filament  the main kinetic channels counteracting electron decay are associative ionization (R190), detachment with \(\text{O}_2\) collisions (R91), and detachment with \(\text{O}_2^-\) collisions (R86). The primary kinetic process for electron consumption is dissociative recombination (R69), which dominates initially. However, associative ionization and detachment become significant after 10 ns. The heating pulse introduces additional processes for electron production, which include avalanche ionization and photo-detachment. The normalized rates for various electron source terms at the laser intensity of 17.6 GW/cm\textsuperscript{2} are shown in Figure \ref{fig:resCoupleDP}. During the second pulse (heating ns-laser pulse), avalanche ionization and photo-detachment dominate, with minimal contributions from other processes. At lower laser intensities (around 1 GW/cm\textsuperscript{2}, not shown), photo-detachment contributes even more to the production of electrons than avalanche ionization. After the pulse begins, associative ionization and detachment show significant increases in contribution. The reduction in the recombination rate, caused by the high electron temperature, has a negligible effect. Associative ionization and detachment are the dominant channels, slowing the plasma decay after the heating ns-laser pulse.

\begin{figure}[]
    \centering
    \includegraphics[width=\linewidth]{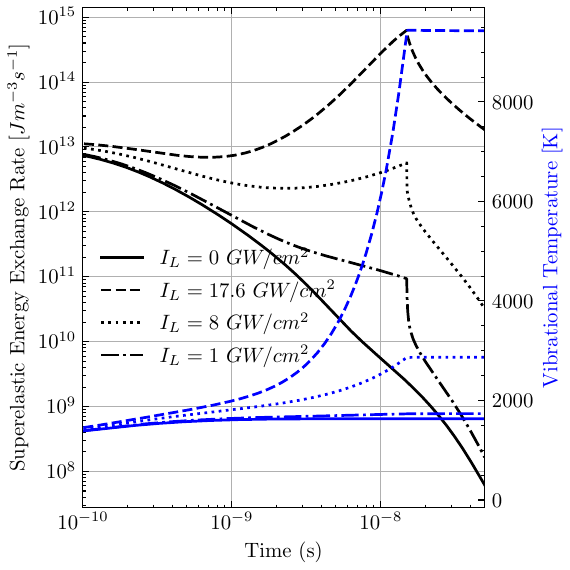}
    \caption{Superelastic energy exchange rate ($Q_{EV}$, left axis) and vibrational temperature (right axis) for a single fs-pulse ($I_L = 0$) and fs-ns laser pulses.}
    \label{fig:reasonCoupTv}
\end{figure}

A critical aspect of energy coupling  by the ns pulse to the fs laser filament is the vibrational excitation of nitrogen molecules due to electron-neutral collisions. After the ns-laser pulse, energy transfer from excited vibrational states to free electrons ($Q_{VE}$) also increases, which helps to sustain relatively high electron temperatures and provides an energy source to the bulk gas through vibrational-translational relaxation. Figure \ref{fig:reasonCoupTv} shows the temporal evolution of vibrational temperature and the source term $Q_{VE}$ at various laser intensities, for cases with and without the second laser pulse (ns-laser ). At lower laser intensities ($\approx$ 1 GW/cm\textsuperscript{2}), the enhancement of the source term $Q_{VE}$ due to inelastic collisions is negligible compared to higher laser intensities. Although the direct influence of the electron temperature on the plasma decay dynamics is minimal \cite{aleksandrov_decay_2016, pokharelSelfconsistentModelNumerical2023a}, the time dependence of electron and vibrational temperatures is important for determining the amount of laser energy stored in vibrational degrees of freedom, an established channel for the subsequent relaxation of energy to the bulk gas.


\subsubsection{Plasma revival by the ns laser pulse}

Experimental observations show that increasing the ns pulse energy to revive high plasma density often leads to stochastic breakdown \cite{polynkinSeededOpticalBreakdown2012a}. However, sustaining laser plasmas at relatively high electron densities, such as $n_\mathrm{e} = 10^{21} \, \text{m}^{-3}$, is essential for applications requiring precise plasma parameters at low gas temperatures while reliably avoiding breakdown.The results presented here on revival of plasma were reported in the AIAA SciTech conference proceedings \cite{pokharelPlasmaEnhancementFsFilament2025}.


\begin{figure*}
    \centering
    \subfloat[Electron density evolution for air-plasma at 1 atm.\label{fig:reviveEle_2}]{%
        \includegraphics[width=0.405\textwidth]{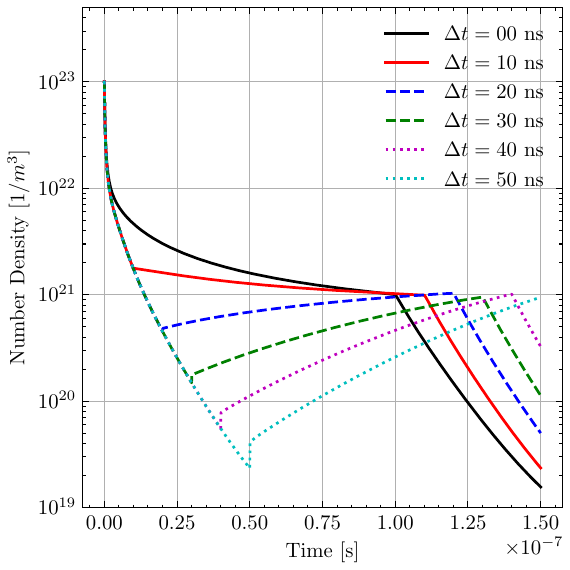}%
    }
    \hspace{0.075\textwidth}%
    \subfloat[Intensity required for the revival.\label{fig:reviveIn_2}]{%
        \includegraphics[width=0.405\textwidth]{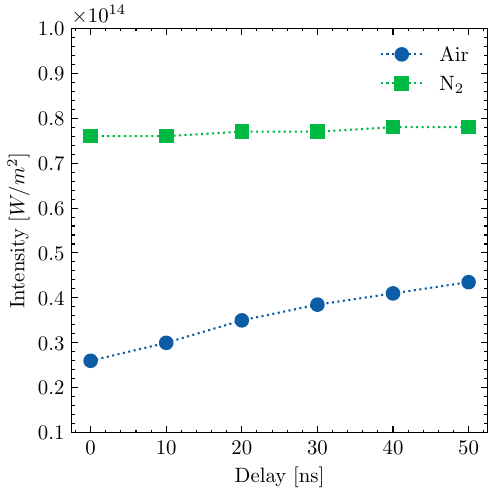}%
    }
    \caption{Revival of the electron density to $\bm{n_\mathrm{e} = 10^{21} \, \text{m}^{-3}}$ 
             in a dual-pulse laser plasma at various time delays of the second ns 
             heating pulse with a pulse width of $\bm{100}$~ns.}
    \label{fig:DpRevival_2}
\end{figure*}

\begin{figure}[h!]
    \centering
    \includegraphics[width=0.8\linewidth]{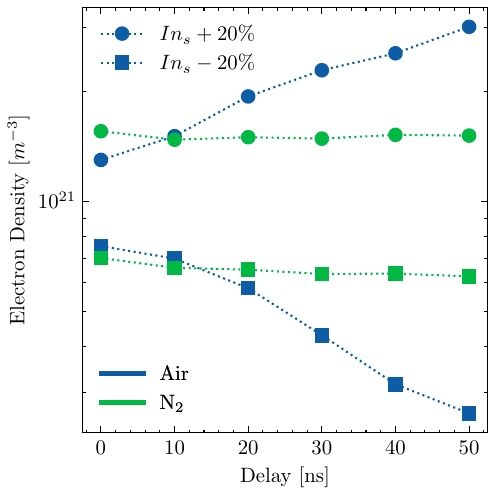}
    \caption{Electron density at the end of the ns-heating pulse (pulse width = 100~ns) with a $\bm{\pm 20\%}$ variation in the laser intensity, which is  required to sustain an electron density of $\bm{n_\mathrm{e} = 10^{21} \, \text{m}^{-3}}$ in a dual-pulse laser configuration. The laser intensity required to sustain the electron density is shown in Fig.~\ref{fig:reviveIn_2}. }
    \label{fig:densityOffset}
\end{figure}

Figure~\ref{fig:reviveEle_2} illustrates the time evolution of electron density following laser energy deposition by fs and ns laser pulses in atmospheric pressure air. Simulation results for various time delays between fs and ns pulses, ranging from $0$~ns to $50$~ns, indicate that maintaining a constant electron density requires a slight increase in ns pulse intensity in air. In nitrogen, the laser intensity needed to sustain $n_\mathrm{e} = 1.0 \times 10^{21} \, \text{m}^{-3}$ remains relatively independent of the time delay between pulses, but the required ns pulse energy is higher than in air due to faster recombination channels, particularly $\text{N}_4^+$ recombination, which dominates at early times. As electron density decreases, the influence of the $\text{N}_4^+$ recombination channel diminishes. In air, the lower ionization potential of oxygen and the presence of $\text{O}_2^-$ facilitate photo-detachment under the ns heating pulse, reducing laser energy requirements compared to plasma revival in nitrogen.


Laser intensity variations of \(\pm 20\%\) are incorporated into the simulations to model realistic fluctuations observed in experiments. The base laser intensity required to sustain the desired plasma density, shown in Fig.~\ref{fig:reviveEle_2}, serves as a reference. Figure~\ref{fig:densityOffset} presents the electron density at the end of the ns pulse, indicating that in nitrogen plasmas, electron density variations remain largely independent of the time delay between pulses. In air plasmas, longer time delays lead to greater deviations from the base electron density for the same intensity variations. At higher laser intensities, the electron density remains within \(n_\mathrm{e} \sim 3.0 \times 10^{21} \, \text{m}^{-3}\), while at lower intensities, similar electron densities are maintained for time delays up to $20$~ns, with a more pronounced drop at longer delays, though still within an order of magnitude of the base case. A 20\% increase in laser intensity results in only a slight rise in electron density, reducing the risk of uncontrollable breakdown. Earlier experimental work \cite{pokharelPlasmaEnhancementFsFilament2025} demonstrated a repeatable and reliable revival of plasma in nitrogen using ns-laser pulse heating without transition to breakdown. The results showed that peak electron density was reached with minimal dependency on the delay, consistent with the observations in this study. These results demonstrate the effectiveness of the dual-pulse laser approach in sustaining plasmas at the desired electron density below the breakdown threshold.

\subsubsection{Temporal delay and shaping of ns laser pulse}


\begin{figure*}
    \centering
    \subfloat[Time averaged electron density.\label{fig:intNe_gauss0}]{%
        \includegraphics[width=0.405\textwidth]{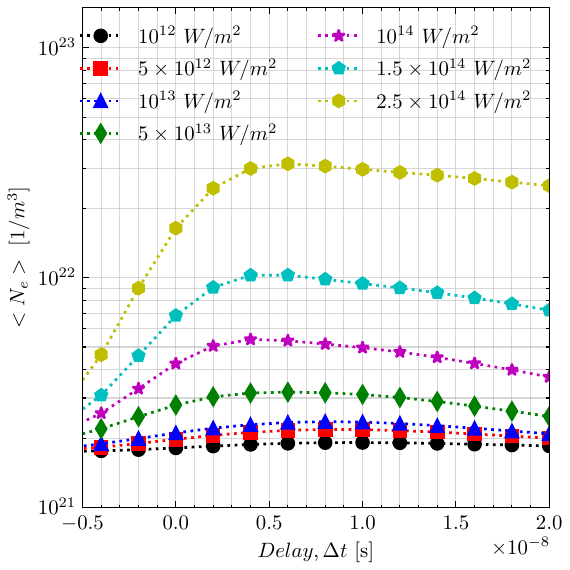}%
    }
    \hspace{0.075\textwidth}%
    \subfloat[Peak gas temperature after the ns laser pulse.\label{fig:thermal_gauss0}]{%
        \includegraphics[width=0.405\textwidth]{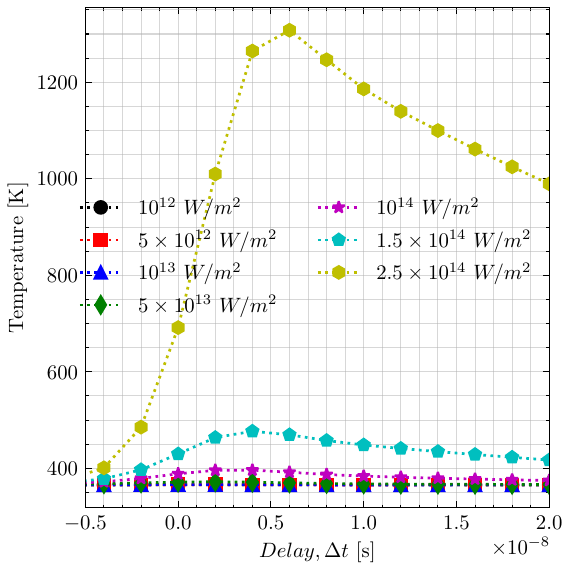}%
    }
    \caption{Plasma revival for a range of intensities of the ns laser pulse: 
             the average electron density (left) and gas temperature (right) 
             for a range of time delays between pulses.}
    \label{fig:DP_delay_gauss}
\end{figure*}

\begin{figure}[]
    \centering
    \includegraphics[width=0.65\linewidth]{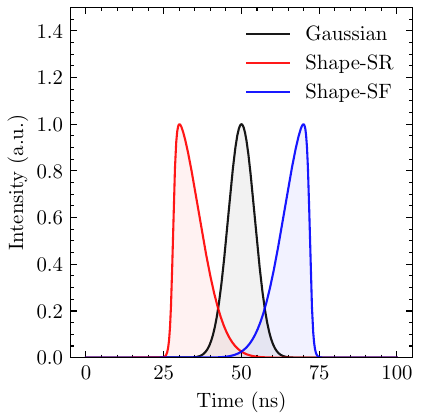}
    \caption{Various temporal shapes of the nanosecond laser pulse.}
    \label{fig:shapes}
\end{figure}


\begin{figure*}
    \centering
    \subfloat[Time averaged electron density.\label{fig:intNe_shape}]{%
        \includegraphics[width=0.405\textwidth]{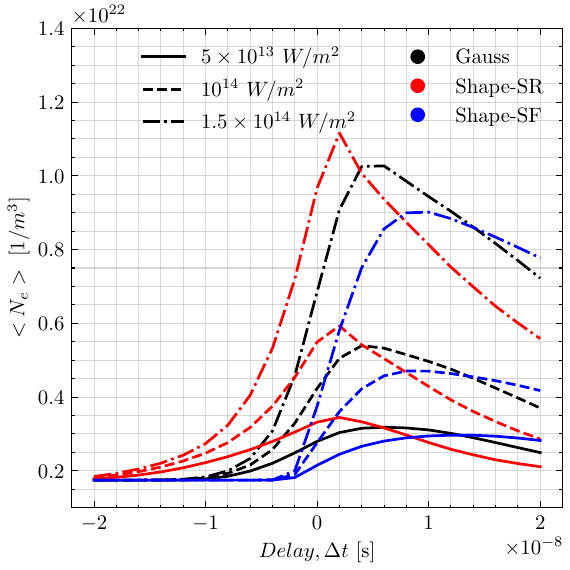}%
    }%
    \hspace{0.075\textwidth}%
    \subfloat[Peak gas temperature after heating by the ns laser pulse.\label{fig:thermal_shape}]{%
        \includegraphics[width=0.405\textwidth]{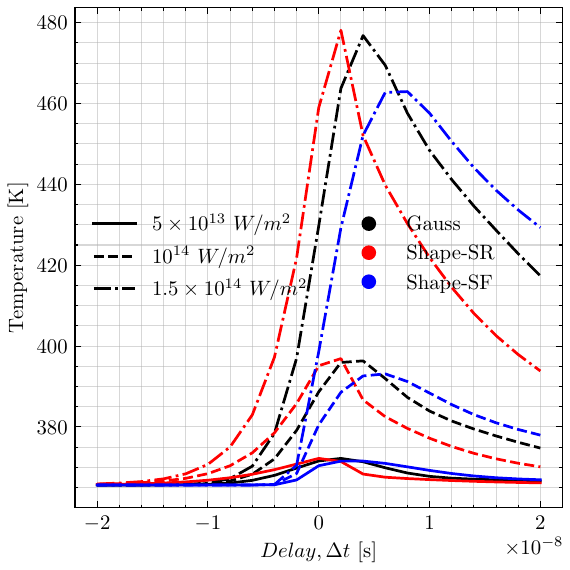}%
    }
    \caption{Comparison of performance for temporally shaped ns laser pulses in 
             dual-pulse configuration: the average electron density (left) and 
             gas temperature (right) against time delay between pulses at various 
             laser intensities of the shaped nanosecond pulse (shapes shown in 
             Eq. \ref{eq:shapePulse}).}
    \label{fig:DP_delay_shape}
\end{figure*}

Finally, we concentrate on the optimization of the dual-pulse laser energy deposition and overall performance of the fs-ns laser pulses in creating and sustaining laser plasmas. 
Preliminary work on temporal delay and shaping of the ns-laser heating pulse was presented earlier \cite{bakLaserIntensityShaping2024}. For consistent comparison of the second ns pulse's performance following preionization from the fs pulse, various metrics such as a peak electron density and a plasma decay time can be considered. However, the strong dependence of the electron density on the time delay between pulses and the laser intensity complicates the establishment of a universal comparison metric. We propose the electron density averaged over a sufficiently long period of time, \(\tau \approx 50~\text{ns} > \Delta t\) as the metric for current analysis. This metric accounts for both the peak electron density and the decay timescale. Furthermore, temporal integration of the electron density reflects energy absorption, as fast gas heating processes, such as electron impact dissociation and dissociative recombination, scale linearly with electron density. Additionally, the peak gas temperature after the heating pulse serves as a direct measure of energy absorption, analogous to the average electron density. In simulations both laser beams are Gaussian in time, the initial pre-ionized channel is created by the fs pulse, with an electron density of \(n_e = 10^{23}~\text{m}^{-3}\) and \(T_e = 3~\text{eV}\) in dry air. The ns heating pulse is a 1064 nm wavelength laser with a 10 ns pulse-width.

The simulation results for the average electron density, \(\langle n_e \rangle = \tau^{-1} \int_0^{\tau} n_e(t) \, dt\), and peak gas temperature for various time delays between fs and the ns pulses at different intensities of the ns pulse are presented in Fig.~\ref{fig:DP_delay_gauss}. We observe that the higher laser intensities result in higher average electron density and stronger thermalization. At an intensity of 25 GW/cm\textsuperscript{2}, the gas temperature rises from about 300 K to 1300 K due to heating from the second ns laser pulse at the optimal time delay. Using the average electron density as a comparison metric, optimal coupling/pairing occurs around half the pulse width time delay for moderate laser intensities (\(I_n >  10^{14}\) W/m\textsuperscript{2}). At lower laser intensities (\(I_n <  10^{14}\) W/m\textsuperscript{2}), optimal pairing shifts to longer time delays between pulses. Similar results were obtained for a nanosecond pulse duration of 16 ns. Consistent with the present findings, earlier experimental work \cite{bakLaserIntensityShaping2024} has shown that optimal energy coupling occurs at a positive offset. Photo-detachment is the primary coupling mechanism at lower laser intensities, with the peak of \Otwm concentration occurring at about 10 ns. The analysis based on thermalization, indicated by the gas temperature, aligns well with the optimal time delay calculated from the proposed metric at moderate laser intensities. However, weak thermalization at lower laser intensities limits the effectiveness of the thermalization-based evaluation of optimal coupling due to the minimal temperature increase.

The advanced capabilities in optics enable the generation of non-Gaussian pulses in time and space. Two shapes, sharp-rise (shape-SR) and sharp-fall (shape-SF), in time are considered for the ns heating pulse to evaluate effects of temporal shaping of the pulse. The non-Gaussian pulse shape is represented by Eq. \ref{eq:shapePulse}, normalized by the maximum amplitude and the pulse width to keep the same energy as the Gaussian laser pulse for accurate comparison.

\begin{align}
I_L(t) &= \tilde{A} \frac{1}{2} \left(1 - \beta \tanh\left(\frac{t - D - \beta \epsilon }{\tau_f}\right)\right) \notag \\
&\quad \cdot \exp\left(-\frac{1}{2}\left(\frac{t - D - \beta \epsilon}{\tau_s}\right)^2\right), \quad \beta \in \{-1, 1\} \label{eq:shapePulse}
\end{align}

Here, $\beta$ represents the shape parameter, $D$ is the time delay, $\epsilon$ denotes a temporal shift, $\tau_f$ is the fast time constant, $\tau_s$ is the slow time constant, and $\tilde{A}$ is the normalized laser amplitude. In this case, $\epsilon = 2$ ns and $\tau_f = 1$ ns, with other parameters scaled to ensure same energy as the Gaussian pulse. The profiles of these pulses are shown in Fig.~\ref{fig:shapes}. The time delay is defined as the time difference between the fs pulse and the peak of the shaped pulse.

For analysis, the plasma generated by  the fs pulse is maintained at 0 ns while the second ns pulse is applied with the time delay ranging from -20 ns to 20 ns to evaluate the performance metric. The time-averaged electron density and bulk gas temperature are shown in Fig. \ref{fig:DP_delay_shape} for different laser intensities of the ns-laser pulse.  Optimal coupling occurs when the electron decay profile closely correlates with the temporal profile of the ns-laser intensity, as absorption is directly proportional to the product of the electron density and laser intensity. Consequently, the laser intensity shape-SR outperforms other pulses due to its stronger correlation with the electron decay curve. The laser intensity shape-SR is preferable for the short time delay between pulses, while the shape-SF is favored for the longer time delay, with the Gaussian pulse performing better in intermediate scenarios. Moreover, for the longer time delays, the earlier tail of the shape-SF pulse experiences higher electron density regions, explaining the improved performance even for longer time delay between pulses. Thus, the shaped laser pulses demonstrate enhanced performance at corresponding optimal temporal delay conditions compared to the Gaussian.

\section{Conclusions}

The properties of laser-induced plasma have been extensively studied for applications in engineering, manufacturing, diagnostics, and defense. Accurate and reliable methodologies and models are essential for bridging the gap between laboratory studies and real-world applications. However, understanding the underlying physics and developing predictive models for quantitative comparisons remain significant challenges. While recent experimental efforts have explored additional parameters, computational studies providing detailed analyses of the temporal and spatial dynamics of various species and temperatures remain limited.

This work extends a previously developed \(\text{N}_2\) plasma kinetics model to enable detailed modeling of non-equilibrium laser plasmas in \(\text{N}_2\)-\(\text{O}_2\) mixtures. Using a self-consistent approach, the model incorporates enhanced kinetics, including associative ionization and energy coupling mechanisms such as avalanche ionization and photodetachment. It is integrated with the multidimensional plasma solver LOTASFOAM \cite{pokharelSelfconsistentModelNumerical2023a}. Simulation results are compared with available experimental data from the literature, showing reasonable agreement in electron dynamics, gas temperature, and refractive index.

After model validation, the plasma dynamics of fs-laser-induced filaments in atmospheric-pressure air are examined and compared with laser plasmas in pure nitrogen. Quantitative characteristics of plasma properties for fs-laser filament dynamics are provided, covering a range of initial electron densities and electron temperatures. Additionally, fs-ns dual-pulse laser plasma properties are studied to understand and optimize the energy coupling of the heating ns-laser pulse to the fs-laser filament. Additional experimental measurements of vibrational temperatures, atomic species concentrations, and gas temperature dynamics would be valuable for comparison with the computational results and for enhancing the understanding of non-equilibrium plasmas and their interaction with laser pulses. The main findings are summarized as follows: a) A small addition of \(\text{O}_2\) to the nitrogen-oxygen mixture results in a slower decay of electron density; b) In air plasma, the key difference compared to plasma decay in pure nitrogen is the suppression of the N\(_4^{+}\) production and recombination channel at early times due to the presence of \(\text{O}_2\); c) An accumulation of negatively charged O\(_2^-\) ions is observed at the radial edges of the plasma channel; d) A comparison of single ns-pulse and dual-pulse (fs-ns) plasma reveals improved decay characteristics and reduced sensitivity to small variations in laser parameters, enabling repeatable revival of plasma to an electron density of \( n_e = 10^{21}~\text{m}^{-3} \); e) Associative ionization and detachment are key kinetic channels that influence temporal plasma dynamics following the ns-heating pulse in dual-pulse scenarios; f) Optimal energy coupling between the fs-laser filament and heating ns-laser pulse occurs at the delay of half the pulse width for Gaussian pulses; g) In fs-ns dual pulse, the shaped ns pulses outperform Gaussian pulses at specific time windows.

\begin{acknowledgments}

This work was supported by the Office of Naval Research under grant N00014-22-1-2235 (Program Manager, Mr. Quentin Saulter). 
The author, S.P., would like to thank Dr. Richard Miles and Dr. Arthur Dogariu for their valuable discussions. A portion of this research was carried out using the advanced computing resources of TERRA and GRACE clusters provided by Texas A\&M High-Performance Research Computing.
\end{acknowledgments}

\section*{Data Availability Statement}
The data that support the findings of this study are available from the corresponding author upon reasonable request.

\appendix*
\section{Reaction Sets}
\begin{widetext}

\begin{longtable}{p{1.25cm}p{5.5cm}p{6.0cm}p{2.0cm}p{2.0cm}}
\caption{\label{tab:mytable}Plasma kinetics model for \(\text{N}_2\)-\(\text{O}_2\).} \\
\toprule
\textbf{ID} & \textbf{Reaction} & \textbf{Rate Expression} & \textbf{Units} & \textbf{References} \\ 
\midrule
\endfirsthead
\caption[]{(Continued)} \\
\toprule
\textbf{ID} & \textbf{Reaction} & \textbf{Rate Expression} & \textbf{Units} & \textbf{References} \\ 
\midrule
\endhead


R1 & $2 N + N_2 \rightarrow N_2 + N_2(A)$ & $1.7 \cdot 10^{-33} \cdot 10^{-12}$ & $m^6s^{-1}$ & \cite{obrusnik_electric_2018, pancheshnyi_zdplaskin_2008} \\ 
R2 & $2 N + O_2 \rightarrow N_2(A) + O_2$ & $1.7 \cdot 10^{-33} \cdot 10^{-12}$ & $m^6s^{-1}$ & \cite{obrusnik_electric_2018, pancheshnyi_zdplaskin_2008} \\ 
R3 & $2 N + NO \rightarrow N_2(A) + NO$ & $1.7 \cdot 10^{-33} \cdot 10^{-12}$ & $m^6s^{-1}$ & \cite{obrusnik_electric_2018, pancheshnyi_zdplaskin_2008} \\ 
R4 & $3 N \rightarrow N + N_2(A)$ & $10^{-32} \cdot 10^{-12}$ & $m^6s^{-1}$ & \cite{obrusnik_electric_2018, pancheshnyi_zdplaskin_2008} \\ 
R5 & $2 N + O \rightarrow N_2(A) + O$ & $10^{-32} \cdot 10^{-12}$ & $m^6s^{-1}$ & \cite{obrusnik_electric_2018, pancheshnyi_zdplaskin_2008} \\ 
R6 & $2 N + N_2 \rightarrow N_2 + N_2(B)$ & $2.4 \cdot 10^{-33} \cdot 10^{-12}$ & $m^6s^{-1}$ & \cite{obrusnik_electric_2018, pancheshnyi_zdplaskin_2008} \\ 
R7 & $2 N + O_2 \rightarrow N_2(B) + O_2$ & $2.4 \cdot 10^{-33} \cdot 10^{-12}$ & $m^6s^{-1}$ & \cite{obrusnik_electric_2018, pancheshnyi_zdplaskin_2008} \\ 
R8 & $2 N + NO \rightarrow N_2(B) + NO$ & $2.4 \cdot 10^{-33} \cdot 10^{-12}$ & $m^6s^{-1}$ & \cite{obrusnik_electric_2018, pancheshnyi_zdplaskin_2008} \\ 
R9 & $3 N \rightarrow N + N_2(B)$ & $1.4 \cdot 10^{-32} \cdot 10^{-12}$ & $m^6s^{-1}$ & \cite{obrusnik_electric_2018, pancheshnyi_zdplaskin_2008} \\ 
R10 & $2 N + O \rightarrow N_2(B) + O$ & $1.4 \cdot 10^{-32} \cdot 10^{-12}$ & $m^6s^{-1}$ & \cite{obrusnik_electric_2018, pancheshnyi_zdplaskin_2008} \\ 
R11 & $N_2(A) \rightarrow N_2 + h\nu$ & $0.5$ & $s^{-1}$ & \cite{peters_considerations_2019, shneider_population_2011, zhang_enhancement_2016} \\ 
R12 & $N_2(B) \rightarrow N_2(A) + h\nu$ & $152000.0$ & $s^{-1}$ & \cite{peters_considerations_2019, shneider_population_2011, zhang_enhancement_2016} \\ 
R13 & $N_2(C) \rightarrow N_2(B) + h\nu$ & $26900000.0$ & $s^{-1}$ & \cite{peters_considerations_2019, shneider_population_2011, zhang_enhancement_2016} \\ 
R14 & $2 N_2(A) \rightarrow N_2(v) + N_2(B)$ & $2.9 \cdot 10^{-15} \sqrt{\frac{T}{300}}$ & $m^3s^{-1}$ & \cite{peters_considerations_2019, shneider_population_2011, zhang_enhancement_2016} \\ 
R15 & $2 N_2(A) \rightarrow N_2(v) + N_2(C)$ & $2.6 \cdot 10^{-16} \sqrt{\frac{T}{300}}$ & $m^3s^{-1}$ & \cite{peters_considerations_2019, shneider_population_2011, zhang_enhancement_2016} \\ 
R16 & $N_4^+ + e^- \rightarrow N_2 + N_2(A)$ & $0.02 \cdot 1.4 \cdot 10^{-12} \left(\frac{300.0}{T_{e}}\right)^{0.41}$ & $m^3s^{-1}$ & \cite{peters_considerations_2019, shneider_population_2011, zhang_enhancement_2016} \\ 
R17 & $N_4^+ + e^- \rightarrow N_2 + N_2(B)$ & $0.87 \cdot 1.4 \cdot 10^{-12} \left(\frac{300.0}{T_{e}}\right)^{0.41}$ & $m^3s^{-1}$ & \cite{peters_considerations_2019, shneider_population_2011, zhang_enhancement_2016} \\ 
R18 & $N_4^+ + e^- \rightarrow N_2 + N_2(C)$ & $0.11 \cdot 1.4 \cdot 10^{-12} \left(\frac{300.0}{T_{e}}\right)^{0.41}$ & $m^3s^{-1}$ & \cite{peters_considerations_2019, shneider_population_2011, zhang_enhancement_2016} \\ 
R19 & $N_4^+ + 2 e^- \rightarrow 2 N_2 + e^-$ & $7.0 \cdot 10^{-32} \left(\frac{300}{T_{e}}\right)^{4.5}$ & $m^6s^{-1}$ & \cite{peters_considerations_2019, shneider_population_2011, zhang_enhancement_2016} \\ 
R20 & $N_3^+ + e^- \rightarrow N + N_2$ & $2.0 \cdot 10^{-13} \sqrt{\frac{300}{T_{e}}}$ & $m^3s^{-1}$ & \cite{peters_considerations_2019, shneider_population_2011, zhang_enhancement_2016} \\ 
R21 & $N_2^+ + e^- \rightarrow N + N(D)$ & $1.8 \cdot 10^{-13} \left(\frac{300.0}{T_{e}}\right)^{0.39} \cdot 0.46$ & $m^3s^{-1}$ & \cite{peters_considerations_2019, shneider_population_2011, zhang_enhancement_2016} \\ 
R22 & $N_2^+ + e^- \rightarrow 2 N(D)$ & $1.8 \cdot 10^{-13} \left(\frac{300.0}{T_{e}}\right)^{0.39} \cdot 0.46$ & $m^3s^{-1}$ & \cite{peters_considerations_2019, shneider_population_2011, zhang_enhancement_2016} \\ 
R23 & $N_2^+ + e^- \rightarrow N + N(P)$ & $1.8 \cdot 10^{-13} \left(\frac{300.0}{T_{e}}\right)^{0.39} \cdot 0.08$ & $m^3s^{-1}$ & \cite{peters_considerations_2019, shneider_population_2011, zhang_enhancement_2016} \\ 
R24 & $N_2^+ + e^- \rightarrow N_2 + h\nu$ & $4.0 \cdot 10^{-18} \left(\frac{300}{T_{e}}\right)^{0.7}$ & $m^3s^{-1}$ & \cite{peters_considerations_2019, shneider_population_2011, zhang_enhancement_2016} \\ 
R25 & $N_2 + N_2^+ + e^- \rightarrow 2 N_2$ & $6.0 \cdot 10^{-39} \left(\frac{300}{T_{e}}\right)^{1.5}$ & $m^6s^{-1}$ & \cite{peters_considerations_2019, shneider_population_2011, zhang_enhancement_2016} \\ 
R26 & $N_2^+ + 2 e^- \rightarrow N_2 + e^-$ & $2.0 \cdot 10^{-31} \left(\frac{300}{T_{e}}\right)^{4.5}$ & $m^6s^{-1}$ & \cite{peters_considerations_2019, shneider_population_2011, zhang_enhancement_2016} \\ 
R27 & $N_2 + e^- \rightarrow N_2^+ + 2 e^-$ & $\frac{5.05 \cdot 10^{-17} \left(\sqrt{T_{e}} + 1.1 \cdot 10^{-5} T_{e}^{1.5}\right)}{e^{\frac{182000.0}{T_{e}}}}$ & $m^3s^{-1}$ & \cite{peters_considerations_2019, shneider_population_2011, zhang_enhancement_2016} \\ 
R28 & $N^+ + e^- \rightarrow N$ & $3.5 \cdot 10^{-18} \left(\frac{300}{T_{e}}\right)^{0.7}$ & $m^3s^{-1}$ & \cite{peters_considerations_2019, shneider_population_2011, zhang_enhancement_2016} \\ 
R29 & $N_2 + N^+ + e^- \rightarrow N + N_2$ & $6.0 \cdot 10^{-39} \left(\frac{300}{T_{e}}\right)^{1.5}$ & $m^6s^{-1}$ & \cite{peters_considerations_2019, shneider_population_2011, zhang_enhancement_2016} \\ 
R30 & $N^+ + 2 e^- \rightarrow N + e^-$ & $2.0 \cdot 10^{-31} \left(\frac{300}{T_{e}}\right)^{4.5}$ & $m^6s^{-1}$ & \cite{peters_considerations_2019, shneider_population_2011, zhang_enhancement_2016} \\ 
R31 & $N_2(A) + N_3^+ \rightarrow 2 N_2 + N^+$ & $6.0 \cdot 10^{-16}$ & $m^3s^{-1}$ & \cite{peters_considerations_2019, shneider_population_2011, zhang_enhancement_2016} \\ 
R32 & $N_2(A) + N_2^+ \rightarrow N + N_3^+$ & $3.0 \cdot 10^{-16}$ & $m^3s^{-1}$ & \cite{peters_considerations_2019, shneider_population_2011, zhang_enhancement_2016} \\ 
R33 & $N_2(A) + N_2^+ \rightarrow N + N_2 + N^+$ & $4.0 \cdot 10^{-16}$ & $m^3s^{-1}$ & \cite{peters_considerations_2019, shneider_population_2011, zhang_enhancement_2016} \\ 
R34 & $N_2 + N_2(A) \rightarrow 2 N_2$ & $2.0 \cdot 10^{-23}$ & $m^3s^{-1}$ & \cite{peters_considerations_2019, shneider_population_2011, zhang_enhancement_2016} \\ 
R35 & $N + N_2(A) \rightarrow N + N_2$ & $6.2 \cdot 10^{-17} \left(\frac{300}{T}\right)^{2/3}$ & $m^3s^{-1}$ & \cite{peters_considerations_2019, shneider_population_2011, zhang_enhancement_2016} \\ 
R36 & $N_2(v) + N_2(B) \rightarrow N_2(v) + N_2(A)$ & $1.2 \cdot 10^{-17}$ & $m^3s^{-1}$ & \cite{peters_considerations_2019, shneider_population_2011, zhang_enhancement_2016} \\ 
R37 & $N_2(v) + N_2(C) \rightarrow N_2(v) + N_2(B)$ & $1.2 \cdot 10^{-17} \left(\frac{300}{T}\right)^{0.33}$ & $m^3s^{-1}$ & \cite{peters_considerations_2019, shneider_population_2011, zhang_enhancement_2016} \\ 
R38 & $N_2 + N_4^+ \rightarrow 2 N_2 + N_2^+$ & $2.1 \cdot 10^{-22} e^{\frac{T}{121}}$ & $m^3s^{-1}$ & \cite{peters_considerations_2019, shneider_population_2011, zhang_enhancement_2016} \\ 
R39 & $N + N_4^+ \rightarrow 2 N_2 + N^+$ & $10^{-17}$ & $m^3s^{-1}$ & \cite{peters_considerations_2019, shneider_population_2011, zhang_enhancement_2016} \\ 
R40 & $N + N_4^+ \rightarrow N_2 + N_3^+$ & $10^{-15}$ & $m^3s^{-1}$ & \cite{peters_considerations_2019, shneider_population_2011, zhang_enhancement_2016} \\ 
R41 & $N + N_3^+ \rightarrow N_2 + N_2^+$ & $6.6 \cdot 10^{-17}$ & $m^3s^{-1}$ & \cite{peters_considerations_2019, shneider_population_2011, zhang_enhancement_2016} \\ 
R42 & $N_2 + N_3^+ \rightarrow 2 N_2 + N^+$ & $6.0 \cdot 10^{-16}$ & $m^3s^{-1}$ & \cite{peters_considerations_2019, shneider_population_2011, zhang_enhancement_2016} \\ 
R43 & $N_2 + N_2^+ \rightarrow N + N_2 + N^+$ & $1.2 \cdot 10^{-17}$ & $m^3s^{-1}$ & \cite{peters_considerations_2019, shneider_population_2011, zhang_enhancement_2016} \\ 
R44 & $N_2 + N_2^+ \rightarrow N + N_3^+$ & $5.5 \cdot 10^{-18}$ & $m^3s^{-1}$ & \cite{peters_considerations_2019, shneider_population_2011, zhang_enhancement_2016} \\ 
R45 & $N + N_2^+ \rightarrow N_2 + N^+$ & $7.2 \cdot 10^{-19} e^{\frac{300}{T}}$ & $m^3s^{-1}$ & \cite{peters_considerations_2019, shneider_population_2011, zhang_enhancement_2016} \\ 
R46 & $2 N_2 + N_2^+ \rightarrow N_4^+ + N_2$ & $4.086 \cdot 10^{-42} \cdot \left(8.16 e^{\frac{-T}{224}} + 1.1 e^{\frac{- T}{997}}\right) $ & $m^6s^{-1}$ & \cite{peters_considerations_2019, shneider_population_2011, zhang_enhancement_2016, troe_temperature_2005,ilyin_emission_2022} \\ 
R47 & $N + N_2 + N_2^+ \rightarrow N_2 + N_3^+$ & $9.0 \cdot 10^{-42} e^{\frac{400}{T}}$ & $m^6s^{-1}$ & \cite{peters_considerations_2019, shneider_population_2011, zhang_enhancement_2016} \\ 
R48 & $N_2 + N^+ \rightarrow N + N_2^+$ & $10^{-19}$ & $m^3s^{-1}$ & \cite{peters_considerations_2019, shneider_population_2011, zhang_enhancement_2016} \\ 
R49 & $2 N_2 + N^+ \rightarrow N_2 + N_3^+$ & $2.0 \cdot 10^{-41} \left(\frac{300}{T}\right)^{2.0}$ & $m^6s^{-1}$ & \cite{peters_considerations_2019, shneider_population_2011, zhang_enhancement_2016} \\ 
R50 & $N + N_2 + N^+ \rightarrow N_2 + N_2^+$ & $10^{-41} \cdot 300 \cdot \frac{1}{T}$ & $m^6s^{-1}$ & \cite{peters_considerations_2019, shneider_population_2011, zhang_enhancement_2016} \\ 
R51 & $N_2 + e^- \rightarrow N + N(D) + e^-$ & $10^{-6} \cdot 4.95 \cdot 10^{-8} \left(\frac{T_{e}}{6000.0}\right)^{0.388} e^{- \frac{113729.0}{T_{e}}}$ & $m^3s^{-1}$ & \cite{peters_considerations_2019, shneider_population_2011, zhang_enhancement_2016, park_rate_2008, niu_assessment_2018, kim_modification_2021} \\ 
R52 & $N + N_2(A) \rightarrow N_2 + N(P)$ & $10^{-6} \cdot 5.0 \cdot 10^{-11}$ & $m^3s^{-1}$ & \cite{peters_considerations_2019, shneider_population_2011, zhang_enhancement_2016, popov_associative_2009, leonov_femtosecond_2012, volynets_n2_2018} \\ 
R53 & $N_2 + e^- \rightarrow N_2(A,v=0-4) + e^-$ & $\mathrm{BOLSIG}$ & $m^3s^{-1}$ & \cite{noauthor_siglo_nodate} \\ 
R54 & $N_2(A,v=0-4) + e^- \rightarrow N_2 + e^-$ & $\mathrm{BOLSIG}$ & $m^3s^{-1}$ & \cite{noauthor_siglo_nodate} \\ 
R55 & $N_2 + e^- \rightarrow N_2(A,v=5-9) + e^-$ & $\mathrm{BOLSIG}$ & $m^3s^{-1}$ & \cite{noauthor_siglo_nodate} \\ 
R56 & $N_2(A,v=5-9) + e^- \rightarrow N_2 + e^-$ & $\mathrm{BOLSIG}$ & $m^3s^{-1}$ & \cite{noauthor_siglo_nodate} \\ 
R57 & $N_2 + e^- \rightarrow N_2(B) + e^-$ & $\mathrm{BOLSIG}$ & $m^3s^{-1}$ & \cite{noauthor_siglo_nodate} \\ 
R58 & $N_2(B) + e^- \rightarrow N_2 + e^-$ & $\mathrm{BOLSIG}$ & $m^3s^{-1}$ & \cite{noauthor_siglo_nodate} \\ 
R59 & $N_2 + e^- \rightarrow N_2(A,v>9) + e^-$ & $\mathrm{BOLSIG}$ & $m^3s^{-1}$ & \cite{noauthor_siglo_nodate} \\ 
R60 & $N_2(A,v>9) + e^- \rightarrow N_2 + e^-$ & $\mathrm{BOLSIG}$ & $m^3s^{-1}$ & \cite{noauthor_siglo_nodate} \\ 
R61 & $N_2 + e^- \rightarrow N_2(C) + e^-$ & $\mathrm{BOLSIG}$ & $m^3s^{-1}$ & \cite{noauthor_siglo_nodate} \\ 
R62 & $N_2(C) + e^- \rightarrow N_2 + e^-$ & $\mathrm{BOLSIG}$ & $m^3s^{-1}$ & \cite{noauthor_siglo_nodate} \\ 
R63 & $N_2 + N_2(a') \rightarrow N_2 + N_2(B)$ & $2.8 \cdot 10^{-13} \cdot 10^{-6}$ & $m^3s^{-1}$ & \cite{peters_considerations_2019, shneider_population_2011, zhang_enhancement_2016, bakReducedSetAir2015} \\ 
R64 & $N_2(a') + O_2 \rightarrow N_2 + O + O(D)$ & $2.8 \cdot 10^{-11} \cdot 10^{-6}$ & $m^3s^{-1}$ & \cite{peters_considerations_2019, shneider_population_2011, zhang_enhancement_2016, bakReducedSetAir2015} \\ 
R65 & $N_2 + e^- \rightarrow N_2(a') + e^-$ & $\mathrm{BOLSIG}$ & $m^3s^{-1}$ & \cite{noauthor_siglo_nodate} \\ 
R66 & $O_2 + e^- \rightarrow O_2^+ + 2 e^-$ & $\mathrm{BOLSIG}$ & $m^3s^{-1}$ & \cite{noauthor_siglo_nodate} \\ 
R67 & $N_2(A) + N_2(a') \rightarrow N_4^+ + e^-$ & $1.5 \cdot 10^{-13} \cdot 10^{-6}$ & $m^3s^{-1}$ & \cite{peters_considerations_2019, shneider_population_2011, zhang_enhancement_2016} \\ 
R68 & $2 N_2(a') \rightarrow N_4^+ + e^-$ & $3.0 \cdot 10^{-13} \cdot 10^{-6}$ & $m^3s^{-1}$ & \cite{peters_considerations_2019, shneider_population_2011, zhang_enhancement_2016} \\ 
R69 & $O_2^+ + e^- \rightarrow 2 O$ & $1.95 \cdot 10^{-13} \left(\frac{300.0}{T_{e}}\right)^{0.7}$ & $m^3s^{-1}$ & \cite{Shneider2011TailoringPulse, kossyi_kinetic_1992, popovFastGasHeating2011,bodrovEffectElectricField2013,florescu-mitchell_dissociative_2006} \\ 
R70 & $O_4^+ + e^- \rightarrow 2 O_2$ & $4.2 \cdot 10^{-12} \left(\frac{300.0}{T_{e}}\right)^{0.5}$ & $m^3s^{-1}$ & \cite{Shneider2011TailoringPulse, kossyi_kinetic_1992, popovFastGasHeating2011,bodrovEffectElectricField2013,florescu-mitchell_dissociative_2006} \\ 
R71 & $NO^+ + e^- \rightarrow N + O$ & $4.0 \cdot 10^{-13} \left(\frac{300.0}{T_{e}}\right)^{1.5}$ & $m^3s^{-1}$ & \cite{Shneider2011TailoringPulse, kossyi_kinetic_1992, popovFastGasHeating2011,bodrovEffectElectricField2013,florescu-mitchell_dissociative_2006} \\ 
R72 & $O_2^+ + 2 e^- \rightarrow 2 O + e^-$ & $2.0 \cdot 10^{-19} \left(\frac{T_{e}}{300.0}\right)^{-4.5} \cdot 10^{-12}$ & $m^6s^{-1}$ & \cite{popovFastGasHeating2011,bodrovEffectElectricField2013,florescu-mitchell_dissociative_2006} \\ 
R73 & $O^+ + 2 e^- \rightarrow O + e^-$ & $1.4 \cdot 10^{-8} T_{e}^{-4.5} \cdot 10^{-12}$ & $m^6s^{-1}$ & \cite{poggieNumericalSimulationNanosecondpulse2012} \\ 
R74 & $O_2 + O_2^+ + e^- \rightarrow 2 O_2$ & $3.1 \cdot 10^{-23} T_{e}^{-1.5} \cdot 10^{-12}$ & $m^6s^{-1}$ & \cite{poggieNumericalSimulationNanosecondpulse2012} \\ 
R75 & $N_2 + O_2^+ + e^- \rightarrow N_2 + O_2$ & $3.1 \cdot 10^{-23} T_{e}^{-1.5} \cdot 10^{-12}$ & $m^6s^{-1}$ & \cite{poggieNumericalSimulationNanosecondpulse2012} \\ 
R76 & $N_2^+ + O_2 + e^- \rightarrow N_2 + O_2$ & $3.1 \cdot 10^{-23} T_{e}^{-1.5} \cdot 10^{-12}$ & $m^6s^{-1}$ & \cite{poggieNumericalSimulationNanosecondpulse2012} \\ 
R77 & $N_2 + O^+ + e^- \rightarrow N_2 + O$ & $3.1 \cdot 10^{-23} T_{e}^{-1.5} \cdot 10^{-12}$ & $m^6s^{-1}$ & \cite{poggieNumericalSimulationNanosecondpulse2012} \\ 
R78 & $O_2 + O^+ + e^- \rightarrow O + O_2$ & $3.1 \cdot 10^{-23} T_{e}^{-1.5} \cdot 10^{-12}$ & $m^6s^{-1}$ & \cite{poggieNumericalSimulationNanosecondpulse2012} \\ 
R79 & $2 O_2 + e^- \rightarrow O_2 + O_2^-$ & $1.4 \cdot 10^{-41} \cdot 300 \cdot \frac{1}{T_{e}} e^{- \frac{600}{T}} e^{\frac{700 \left(- T + T_{e}\right)}{T T_{e}}}$ & $m^6s^{-1}$ & \cite{Shneider2011TailoringPulse} \\ 
R80 & $N_2 + O_2 + e^- \rightarrow N_2 + O_2^-$ & $1.07 \cdot 10^{-43} \cdot 300 \cdot \frac{1}{T_{e}} e^{- \frac{70}{T}} e^{\frac{1500 \left(- T + T_{e}\right)}{T T_{e}}}$ & $m^6s^{-1}$ & \cite{Shneider2011TailoringPulse} \\ 
R81 & $O + O_2 + e^- \rightarrow O + O_2^-$ & $10^{-43}$ & $m^6s^{-1}$ & \cite{Shneider2011TailoringPulse} \\ 
R82 & $O + O_2 + e^- \rightarrow O_2 + O^-$ & $10^{-43}$ & $m^6s^{-1}$ & \cite{Shneider2011TailoringPulse} \\ 
R83 & $O_3 + e^- \rightarrow O + O_2^-$ & $1.92 \cdot 10^{-9} \left(\frac{T_{e}}{11600}\right)^{-1.5} e^{\left(-1.3\right) 11600 \cdot \frac{1}{T_{e}}} 10^{-6}$ & $m^3s^{-1}$ & \cite{Shneider2011TailoringPulse, cicman_rate_2003} \\ 
R84 & $O_3 + e^- \rightarrow O_2 + O^-$ & $5.87 \cdot 10^{-9} \left(\frac{T_{e}}{11600}\right)^{-1.5} e^{\left(-1.59\right) 11600 \cdot \frac{1}{T_{e}}} 10^{-6}$ & $m^3s^{-1}$ & \cite{Shneider2011TailoringPulse, cicman_rate_2003} \\ 
R85 & $N_2 + O_2^- \rightarrow N_2 + O_2 + e^-$ & $10^{-6} \cdot 1.9 \cdot 10^{-12} \left(\frac{T}{300}\right)^{0.5} e^{- \frac{4990.0}{T}}$ & $m^3s^{-1}$ & \cite{kossyi_kinetic_1992} \\ 
R86 & $O_2 + O_2^- \rightarrow 2 O_2 + e^-$ & $10^{-6} \cdot 2.7 \cdot 10^{-10} \left(\frac{T}{300}\right)^{0.5} e^{- \frac{5590.0}{T}}$ & $m^3s^{-1}$ & \cite{kossyi_kinetic_1992} \\ 
R87 & $N_2(A) + O_2^- \rightarrow N_2 + O_2 + e^-$ & $10^{-6} \cdot 2.1 \cdot 10^{-9}$ & $m^3s^{-1}$ & \cite{kossyi_kinetic_1992} \\ 
R88 & $N_2(B) + O_2^- \rightarrow N_2 + O_2 + e^-$ & $10^{-6} \cdot 2.5 \cdot 10^{-9}$ & $m^3s^{-1}$ & \cite{kossyi_kinetic_1992} \\ 
R89 & $N_2(A) + O^- \rightarrow N_2 + O + e^-$ & $10^{-6} \cdot 2.2 \cdot 10^{-9}$ & $m^3s^{-1}$ & \cite{kossyi_kinetic_1992} \\ 
R90 & $N_2(B) + O^- \rightarrow N_2 + O + e^-$ & $10^{-6} \cdot 1.9 \cdot 10^{-9}$ & $m^3s^{-1}$ & \cite{kossyi_kinetic_1992} \\ 
R91 & $O + O_2^- \rightarrow O_3 + e^-$ & $10^{-6} \cdot 1.5 \cdot 10^{-10}$ & $m^3s^{-1}$ & \cite{kossyi_kinetic_1992} \\ 
R92 & $N + O^- \rightarrow NO + e^-$ & $10^{-6} \cdot 2.6 \cdot 10^{-10}$ & $m^3s^{-1}$ & \cite{kossyi_kinetic_1992} \\ 
R93 & $O_2 + O^- \rightarrow O_3 + e^-$ & $10^{-6} \cdot 5.0 \cdot 10^{-15}$ & $m^3s^{-1}$ & \cite{kossyi_kinetic_1992} \\ 
R94 & $N_2(A) + O_2 \rightarrow N_2 + 2 O$ & $1.7 \cdot 10^{-12} \cdot 10^{-6}$ & $m^3s^{-1}$ & \cite{popov_associative_2009, leonov_femtosecond_2012, volynets_n2_2018, poggieNumericalSimulationNanosecondpulse2012} \\ 
R95 & $N_2(A) + O_2 \rightarrow N_2 + O_2$ & $7.5 \cdot 10^{-13} \cdot 10^{-6}$ & $m^3s^{-1}$ & \cite{popov_associative_2009, leonov_femtosecond_2012, volynets_n2_2018, poggieNumericalSimulationNanosecondpulse2012} \\ 
R96 & $N_2(A) + O \rightarrow N_2 + O(D)$ & $3.0 \cdot 10^{-11} \cdot 10^{-6}$ & $m^3s^{-1}$ & \cite{popov_associative_2009, leonov_femtosecond_2012, volynets_n2_2018, poggieNumericalSimulationNanosecondpulse2012} \\ 
R97 & $N_2(A) + NO \rightarrow N_2 + NO$ & $2.1 \cdot 10^{-11} \cdot 10^{-6}$ & $m^3s^{-1}$ & \cite{popov_associative_2009, leonov_femtosecond_2012, volynets_n2_2018, poggieNumericalSimulationNanosecondpulse2012} \\ 
R98 & $N_2(A) + O \rightarrow N + NO$ & $7.0 \cdot 10^{-12} \cdot 10^{-6}$ & $m^3s^{-1}$ & \cite{popov_associative_2009, leonov_femtosecond_2012, volynets_n2_2018, poggieNumericalSimulationNanosecondpulse2012} \\ 
R99 & $N_2(B) + O_2 \rightarrow N_2 + 2 O$ & $3.0 \cdot 10^{-10} \cdot 10^{-6}$ & $m^3s^{-1}$ & \cite{popov_associative_2009, leonov_femtosecond_2012, volynets_n2_2018, poggieNumericalSimulationNanosecondpulse2012} \\ 
R100 & $N_2(B) + NO \rightarrow N_2(A) + NO$ & $2.4 \cdot 10^{-10} \cdot 10^{-6}$ & $m^3s^{-1}$ & \cite{popov_associative_2009, leonov_femtosecond_2012, volynets_n2_2018, poggieNumericalSimulationNanosecondpulse2012} \\ 
R101 & $N_2(C) + O_2 \rightarrow N_2 + O + O(D)$ & $2.5 \cdot 10^{-10} \cdot 10^{-6}$ & $m^3s^{-1}$ & \cite{popov_associative_2009, leonov_femtosecond_2012, volynets_n2_2018, poggieNumericalSimulationNanosecondpulse2012} \\ 
R102 & $O_2 + O(D) \rightarrow O + O_2$ & $6.4 \cdot 10^{-12} e^{\frac{67.0}{T}} 10^{-6}$ & $m^3s^{-1}$ & \cite{obrusnik_electric_2018, pancheshnyi_zdplaskin_2008} \\ 
R103 & $N_2 + O(D) \rightarrow N_2 + O$ & $2.3 \cdot 10^{-11} \cdot 10^{-6}$ & $m^3s^{-1}$ & \cite{obrusnik_electric_2018, pancheshnyi_zdplaskin_2008} \\ 
R104 & $O + O(D) \rightarrow 2 O$ & $8.0 \cdot 10^{-12} \cdot 10^{-6}$ & $m^3s^{-1}$ & \cite{obrusnik_electric_2018, pancheshnyi_zdplaskin_2008} \\ 
R105 & $O_3 + O(D) \rightarrow 2 O + O_2$ & $1.2 \cdot 10^{-10} \cdot 10^{-6}$ & $m^3s^{-1}$ & \cite{obrusnik_electric_2018, pancheshnyi_zdplaskin_2008} \\ 
R106 & $O_3 + O(D) \rightarrow 2 O_2$ & $1.2 \cdot 10^{-10} \cdot 10^{-6}$ & $m^3s^{-1}$ & \cite{obrusnik_electric_2018, pancheshnyi_zdplaskin_2008} \\ 
R107 & $NO + O(D) \rightarrow N + O_2$ & $1.7 \cdot 10^{-10} \cdot 10^{-6}$ & $m^3s^{-1}$ & \cite{obrusnik_electric_2018, pancheshnyi_zdplaskin_2008} \\ 
R108 & $O + O_2^- \rightarrow O_2 + O^-$ & $10^{-6} \cdot 3.3 \cdot 10^{-10}$ & $m^3s^{-1}$ & \cite{kossyi_kinetic_1992} \\ 
R109 & $O_2 + O^- \rightarrow O + O_2^-$ & $10^{-6} \cdot 10^{-10}$ & $m^3s^{-1}$ & \cite{kossyi_kinetic_1992} \\ 
R110 & $N_4^+ + O_2 \rightarrow 2 N_2 + O_2^+$ & $10^{-6} \cdot 2.5 \cdot 10^{-10}$ & $m^3s^{-1}$ & \cite{kossyi_kinetic_1992} \\ 
R111 & $N_2^+ + O_2 \rightarrow N_2 + O_2^+$ & $10^{-6} \cdot 6.0 \cdot 10^{-11}$ & $m^3s^{-1}$ & \cite{kossyi_kinetic_1992} \\ 
R112 & $2 O_2 + O_2^+ \rightarrow O_2 + O_4^+$ & $2.4 \cdot 10^{-30} \left(\frac{300.0}{T}\right)^{3.2} \cdot 10^{-12}$ & $m^6s^{-1}$ & \cite{kossyi_kinetic_1992} \\ 
R113 & $N_2 + O_2 + O_2^+ \rightarrow N_2 + O_4^+$ & $3.3 \cdot 10^{-35} T^{-3.2} \cdot 10^{-12}$ & $m^6s^{-1}$ & \cite{kossyi_kinetic_1992} \\ 
R114 & $N_2 + N_2^+ + O_2 \rightarrow N_4^+ + O_2$ & $5.0 \cdot 10^{-29} \cdot 10^{-12}$ & $m^6s^{-1}$ & \cite{kossyi_kinetic_1992} \\ 
R115 & $N_2 + O_4^+ \rightarrow N_2 + O_2 + O_2^+$ & $3.3 \cdot 10^{-12} \left(\frac{300.0}{T}\right)^{4.0} e^{- \frac{5030.0}{T}}$ & $m^3s^{-1}$ & \cite{kossyi_kinetic_1992} \\ 
R116 & $O_2 + O_4^+ \rightarrow 2 O_2 + O_2^+$ & $3.3 \cdot 10^{-12} \left(\frac{300.0}{T}\right)^{4.0} e^{- \frac{5030.0}{T}}$ & $m^3s^{-1}$ & \cite{kossyi_kinetic_1992} \\ 
R117 & $N + O_2 + O^+ \rightarrow NO^+ + O_2$ & $10^{-41}$ & $m^6s^{-1}$ & \cite{kossyi_kinetic_1992} \\ 
R118 & $N + N_2 + O^+ \rightarrow N_2 + NO^+$ & $10^{-41}$ & $m^6s^{-1}$ & \cite{kossyi_kinetic_1992} \\ 
R119 & $N^+ + O_2 \rightarrow N + O_2^+$ & $2.8 \cdot 10^{-16}$ & $m^3s^{-1}$ & \cite{kossyi_kinetic_1992} \\ 
R120 & $N^+ + O_2 \rightarrow NO^+ + O$ & $2.5 \cdot 10^{-16}$ & $m^3s^{-1}$ & \cite{kossyi_kinetic_1992} \\ 
R121 & $N^+ + O \rightarrow N + O^+$ & $10^{-18}$ & $m^3s^{-1}$ & \cite{kossyi_kinetic_1992} \\ 
R122 & $N^+ + O_3 \rightarrow NO^+ + O_2$ & $5.0 \cdot 10^{-16}$ & $m^3s^{-1}$ & \cite{kossyi_kinetic_1992} \\ 
R123 & $NO + N^+ \rightarrow N + NO^+$ & $8.0 \cdot 10^{-16}$ & $m^3s^{-1}$ & \cite{kossyi_kinetic_1992} \\ 
R124 & $NO + N^+ \rightarrow N_2^+ + O$ & $3.0 \cdot 10^{-18}$ & $m^3s^{-1}$ & \cite{kossyi_kinetic_1992} \\ 
R125 & $NO + N^+ \rightarrow N_2 + O^+$ & $10^{-18}$ & $m^3s^{-1}$ & \cite{kossyi_kinetic_1992} \\ 
R126 & $N_2 + O^+ \rightarrow N + NO^+$ & $3.0 \cdot 10^{-18} e^{- 0.00311 T}$ & $m^3s^{-1}$ & \cite{kossyi_kinetic_1992} \\ 
R127 & $N_2^+ + O \rightarrow N + NO^+$ & $1.3 \cdot 10^{-10} \left(\frac{300}{T}\right)^{0.5}$ & $m^3s^{-1}$ & \cite{kossyi_kinetic_1992} \\ 
R128 & $N_2^+ + O \rightarrow N_2 + O^+$ & $10^{-17} \left(\frac{300}{T}\right)^{0.2}$ & $m^3s^{-1}$ & \cite{kossyi_kinetic_1992} \\ 
R129 & $N_2^+ + O_3 \rightarrow N_2 + O + O_2^+$ & $10^{-16}$ & $m^3s^{-1}$ & \cite{kossyi_kinetic_1992} \\ 
R130 & $N_2^+ + NO \rightarrow N_2 + NO^+$ & $3.3 \cdot 10^{-16}$ & $m^3s^{-1}$ & \cite{kossyi_kinetic_1992} \\ 
R131 & $N_2 + O_2^+ \rightarrow NO + NO^+$ & $10^{-23}$ & $m^3s^{-1}$ & \cite{kossyi_kinetic_1992} \\ 
R132 & $N + O_2^+ \rightarrow NO^+ + O$ & $1.2 \cdot 10^{-16}$ & $m^3s^{-1}$ & \cite{kossyi_kinetic_1992} \\ 
R133 & $NO + O_2^+ \rightarrow NO^+ + O_2$ & $4.4 \cdot 10^{-16}$ & $m^3s^{-1}$ & \cite{kossyi_kinetic_1992} \\ 
R134 & $N_4^+ + O \rightarrow 2 N_2 + O^+$ & $2.5 \cdot 10^{-16}$ & $m^3s^{-1}$ & \cite{kossyi_kinetic_1992} \\ 
R135 & $N_4^+ + NO \rightarrow 2 N_2 + NO^+$ & $4.0 \cdot 10^{-16}$ & $m^3s^{-1}$ & \cite{kossyi_kinetic_1992} \\ 
R136 & $O + O_4^+ \rightarrow O_2^+ + O_3$ & $3.0 \cdot 10^{-16}$ & $m^3s^{-1}$ & \cite{kossyi_kinetic_1992} \\ 
R137 & $NO + O_4^+ \rightarrow NO^+ + 2 O_2$ & $10^{-16}$ & $m^3s^{-1}$ & \cite{kossyi_kinetic_1992} \\ 
R138 & $O_2 + O_2^- + O_2^+ \rightarrow 2 O + 2 O_2$ & $10^{-12} \cdot 2.0 \cdot 10^{-25} \left(\frac{300}{T}\right)^{2.5}$ & $m^6s^{-1}$ & \cite{popov_associative_2009, leonov_femtosecond_2012, volynets_n2_2018} \\ 
R139 & $N_2 + O_2^- + O_2^+ \rightarrow N_2 + 2 O + O_2$ & $10^{-12} \cdot 2.0 \cdot 10^{-25} \left(\frac{300}{T}\right)^{2.5}$ & $m^6s^{-1}$ & \cite{popov_associative_2009, leonov_femtosecond_2012, volynets_n2_2018} \\ 
R140 & $O_2 + O_2^- + O_4^+ \rightarrow 2 O + 3 O_2$ & $10^{-12} \cdot 2.0 \cdot 10^{-25} \left(\frac{300}{T}\right)^{2.5}$ & $m^6s^{-1}$ & \cite{popov_associative_2009, leonov_femtosecond_2012, volynets_n2_2018} \\ 
R141 & $N_2 + N_2^+ + O_2^- \rightarrow 2 N_2 + O_2$ & $2.0 \cdot 10^{-25} \left(\frac{300.0}{T}\right)^{2.5} \cdot 10^{-12}$ & $m^6s^{-1}$ & \cite{kossyi_kinetic_1992} \\ 
R142 & $N_2^+ + O_2 + O_2^- \rightarrow N_2 + 2 O_2$ & $2.0 \cdot 10^{-25} \left(\frac{300.0}{T}\right)^{2.5} \cdot 10^{-12}$ & $m^6s^{-1}$ & \cite{kossyi_kinetic_1992} \\ 
R143 & $N_2 + O_2^- + O_2^+ \rightarrow N_2 + 2 O_2$ & $2.0 \cdot 10^{-25} \left(\frac{300.0}{T}\right)^{2.5} \cdot 10^{-12}$ & $m^6s^{-1}$ & \cite{kossyi_kinetic_1992} \\ 
R144 & $O_2 + O_2^- + O_2^+ \rightarrow 3 O_2$ & $2.0 \cdot 10^{-25} \left(\frac{300.0}{T}\right)^{2.5} \cdot 10^{-12}$ & $m^6s^{-1}$ & \cite{kossyi_kinetic_1992} \\ 
R145 & $N_2 + N^+ + O_2^- \rightarrow N + N_2 + O_2$ & $2.0 \cdot 10^{-25} \left(\frac{300.0}{T}\right)^{2.5} \cdot 10^{-12}$ & $m^6s^{-1}$ & \cite{kossyi_kinetic_1992} \\ 
R146 & $N^+ + O_2 + O_2^- \rightarrow N + 2 O_2$ & $2.0 \cdot 10^{-25} \left(\frac{300.0}{T}\right)^{2.5} \cdot 10^{-12}$ & $m^6s^{-1}$ & \cite{kossyi_kinetic_1992} \\ 
R147 & $N_2 + O_2^- + O^+ \rightarrow N_2 + O + O_2$ & $2.0 \cdot 10^{-25} \left(\frac{300.0}{T}\right)^{2.5} \cdot 10^{-12}$ & $m^6s^{-1}$ & \cite{kossyi_kinetic_1992} \\ 
R148 & $O_2 + O_2^- + O^+ \rightarrow O + 2 O_2$ & $2.0 \cdot 10^{-25} \left(\frac{300.0}{T}\right)^{2.5} \cdot 10^{-12}$ & $m^6s^{-1}$ & \cite{kossyi_kinetic_1992} \\ 
R149 & $N_2 + NO^+ + O_2^- \rightarrow N_2 + NO + O_2$ & $2.0 \cdot 10^{-25} \left(\frac{300.0}{T}\right)^{2.5} \cdot 10^{-12}$ & $m^6s^{-1}$ & \cite{kossyi_kinetic_1992} \\ 
R150 & $NO^+ + O_2 + O_2^- \rightarrow NO + 2 O_2$ & $2.0 \cdot 10^{-25} \left(\frac{300.0}{T}\right)^{2.5} \cdot 10^{-12}$ & $m^6s^{-1}$ & \cite{kossyi_kinetic_1992} \\ 
R151 & $N_2 + N_2^+ + O^- \rightarrow 2 N_2 + O$ & $2.0 \cdot 10^{-25} \left(\frac{300.0}{T}\right)^{2.5} \cdot 10^{-12}$ & $m^6s^{-1}$ & \cite{kossyi_kinetic_1992} \\ 
R152 & $N_2^+ + O_2 + O^- \rightarrow N_2 + O + O_2$ & $2.0 \cdot 10^{-25} \left(\frac{300.0}{T}\right)^{2.5} \cdot 10^{-12}$ & $m^6s^{-1}$ & \cite{kossyi_kinetic_1992} \\ 
R153 & $N_2 + O_2^+ + O^- \rightarrow N_2 + O + O_2$ & $2.0 \cdot 10^{-25} \left(\frac{300.0}{T}\right)^{2.5} \cdot 10^{-12}$ & $m^6s^{-1}$ & \cite{kossyi_kinetic_1992} \\ 
R154 & $O_2 + O_2^+ + O^- \rightarrow O + 2 O_2$ & $2.0 \cdot 10^{-25} \left(\frac{300.0}{T}\right)^{2.5} \cdot 10^{-12}$ & $m^6s^{-1}$ & \cite{kossyi_kinetic_1992} \\ 
R155 & $N_2 + N^+ + O^- \rightarrow N + N_2 + O$ & $2.0 \cdot 10^{-25} \left(\frac{300.0}{T}\right)^{2.5} \cdot 10^{-12}$ & $m^6s^{-1}$ & \cite{kossyi_kinetic_1992} \\ 
R156 & $N^+ + O_2 + O^- \rightarrow N + O + O_2$ & $2.0 \cdot 10^{-25} \left(\frac{300.0}{T}\right)^{2.5} \cdot 10^{-12}$ & $m^6s^{-1}$ & \cite{kossyi_kinetic_1992} \\ 
R157 & $N_2 + O^- + O^+ \rightarrow N_2 + 2 O$ & $2.0 \cdot 10^{-25} \left(\frac{300.0}{T}\right)^{2.5} \cdot 10^{-12}$ & $m^6s^{-1}$ & \cite{kossyi_kinetic_1992} \\ 
R158 & $O_2 + O^- + O^+ \rightarrow 2 O + O_2$ & $2.0 \cdot 10^{-25} \left(\frac{300.0}{T}\right)^{2.5} \cdot 10^{-12}$ & $m^6s^{-1}$ & \cite{kossyi_kinetic_1992} \\ 
R159 & $N_2 + NO^+ + O^- \rightarrow N_2 + NO + O$ & $2.0 \cdot 10^{-25} \left(\frac{300.0}{T}\right)^{2.5} \cdot 10^{-12}$ & $m^6s^{-1}$ & \cite{kossyi_kinetic_1992} \\ 
R160 & $NO^+ + O_2 + O^- \rightarrow NO + O + O_2$ & $2.0 \cdot 10^{-25} \left(\frac{300.0}{T}\right)^{2.5} \cdot 10^{-12}$ & $m^6s^{-1}$ & \cite{kossyi_kinetic_1992} \\ 
R161 & $N_2 + N^+ + O^- \rightarrow N_2 + NO$ & $2.0 \cdot 10^{-25} \left(\frac{300.0}{T}\right)^{2.5} \cdot 10^{-12}$ & $m^6s^{-1}$ & \cite{kossyi_kinetic_1992} \\ 
R162 & $N^+ + O_2 + O^- \rightarrow NO + O_2$ & $2.0 \cdot 10^{-25} \left(\frac{300.0}{T}\right)^{2.5} \cdot 10^{-12}$ & $m^6s^{-1}$ & \cite{kossyi_kinetic_1992} \\ 
R163 & $N_2 + O^- + O^+ \rightarrow N_2 + O_2$ & $2.0 \cdot 10^{-25} \left(\frac{300.0}{T}\right)^{2.5} \cdot 10^{-12}$ & $m^6s^{-1}$ & \cite{kossyi_kinetic_1992} \\ 
R164 & $O_2 + O^- + O^+ \rightarrow 2 O_2$ & $2.0 \cdot 10^{-25} \left(\frac{300.0}{T}\right)^{2.5} \cdot 10^{-12}$ & $m^6s^{-1}$ & \cite{kossyi_kinetic_1992} \\ 
R165 & $N + O_2 \rightarrow NO + O$ & $10^{-6} \cdot 1.1 \cdot 10^{-14} T e^{- \frac{3150.0}{T}}$ & $m^3s^{-1}$ & \cite{kossyi_kinetic_1992} \\ 
R166 & $N + O_3 \rightarrow NO + O_2$ & $10^{-6} \cdot 2.0 \cdot 10^{-16}$ & $m^3s^{-1}$ & \cite{kossyi_kinetic_1992} \\ 
R167 & $N + NO \rightarrow N_2 + O$ & $10^{-6} \cdot 1.05 \cdot 10^{-12} T^{0.5}$ & $m^3s^{-1}$ & \cite{kossyi_kinetic_1992} \\ 
R168 & $O + O_3 \rightarrow 2 O_2$ & $10^{-6} \cdot 2.0 \cdot 10^{-11} e^{- \frac{2300.0}{T}}$ & $m^3s^{-1}$ & \cite{kossyi_kinetic_1992} \\ 
R169 & $2 N + O_2 \rightarrow N_2 + O_2$ & $0.5 \cdot 10^{-12} \cdot 8.76 \cdot 10^{-34} e^{\frac{500}{T}}$ & $m^6s^{-1}$ & \cite{kossyi_kinetic_1992} \\ 
R170 & $N_2 + 2 O \rightarrow N_2 + O_2$ & $10^{-12} \cdot 2.76 \cdot 10^{-34} e^{\frac{720.0}{T}}$ & $m^6s^{-1}$ & \cite{kossyi_kinetic_1992} \\ 
R171 & $2 O + O_2 \rightarrow 2 O_2$ & $10^{-12} \cdot 2.45 \cdot 10^{-31} T^{-0.63}$ & $m^6s^{-1}$ & \cite{kossyi_kinetic_1992} \\ 
R172 & $N_2 + O + O_2 \rightarrow N_2 + O_3$ & $\frac{5.6 \cdot 10^{-41}}{T^{2.0}}$ & $m^6s^{-1}$ & \cite{kossyi_kinetic_1992} \\ 
R173 & $O + 2 O_2 \rightarrow O_2 + O_3$ & $\frac{8.6 \cdot 10^{-43}}{T^{1.25}}$ & $m^6s^{-1}$ & \cite{kossyi_kinetic_1992} \\ 
R174 & $N_2 + O \rightarrow N + NO$ & $\frac{1.121 \cdot 10^{-16} e^{- \frac{3.75 \cdot 10^{4}}{T^{1.0}}}}{T^{1.0}}$ & $m^3s^{-1}$ & \cite{kossyi_kinetic_1992} \\ 
R175 & $NO + O \rightarrow N + O_2$ & $5.281 \cdot 10^{-21} T e^{- \frac{1.97 \cdot 10^{4}}{T^{1.0}}}$ & $m^3s^{-1}$ & \cite{kossyi_kinetic_1992} \\ 
R176 & $N + N(P) \rightarrow N + N(D)$ & $1.8 \cdot 10^{-18}$ & $m^3s^{-1}$ & \cite{popov_associative_2009, leonov_femtosecond_2012, volynets_n2_2018} \\ 
R177 & $N(P) + O \rightarrow N(D) + O$ & $10^{-18}$ & $m^3s^{-1}$ & \cite{popov_associative_2009, leonov_femtosecond_2012, volynets_n2_2018} \\ 
R178 & $N(P) + O_2 \rightarrow NO + O(D)$ & $1.25 \cdot 10^{-18}$ & $m^3s^{-1}$ & \cite{popov_associative_2009, leonov_femtosecond_2012, volynets_n2_2018} \\ 
R179 & $N(P) + O_2 \rightarrow NO + O$ & $1.25 \cdot 10^{-18}$ & $m^3s^{-1}$ & \cite{popov_associative_2009, leonov_femtosecond_2012, volynets_n2_2018} \\ 
R180 & $NO + N(P) \rightarrow N_2 + O(D)$ & $1.45 \cdot 10^{-17}$ & $m^3s^{-1}$ & \cite{popov_associative_2009, leonov_femtosecond_2012, volynets_n2_2018} \\ 
R181 & $NO + N(P) \rightarrow N_2 + O$ & $1.45 \cdot 10^{-17}$ & $m^3s^{-1}$ & \cite{popov_associative_2009, leonov_femtosecond_2012, volynets_n2_2018} \\ 
R182 & $N(D) + O_2 \rightarrow NO + O(D)$ & $7.3 \cdot 10^{-18} e^{- \frac{185.0}{T^{1.0}}}$ & $m^3s^{-1}$ & \cite{popov_associative_2009, leonov_femtosecond_2012, volynets_n2_2018} \\ 
R183 & $N(D) + O_2 \rightarrow NO + O$ & $2.4 \cdot 10^{-18} e^{- \frac{185.0}{T^{1.0}}}$ & $m^3s^{-1}$ & \cite{popov_associative_2009, leonov_femtosecond_2012, volynets_n2_2018} \\ 
R184 & $N(D) + O \rightarrow N + O$ & $1.4 \cdot 10^{-18}$ & $m^3s^{-1}$ & \cite{popov_associative_2009, leonov_femtosecond_2012, volynets_n2_2018} \\ 
R185 & $NO + N(D) \rightarrow N_2(v) + O(D)$ & $3.96 \cdot 10^{-17}$ & $m^3s^{-1}$ & \cite{popov_associative_2009, leonov_femtosecond_2012, volynets_n2_2018} \\ 
R186 & $NO + N(D) \rightarrow N_2(v) + O$ & $2.04 \cdot 10^{-17}$ & $m^3s^{-1}$ & \cite{popov_associative_2009, leonov_femtosecond_2012, volynets_n2_2018} \\ 
R187 & $N_2 + N(D) \rightarrow N + N_2$ & $10^{-19} e^{- \frac{510.0}{T^{1.0}}}$ & $m^3s^{-1}$ & \cite{popov_associative_2009, leonov_femtosecond_2012, volynets_n2_2018} \\ 
R188 & $N(D) + N(P) \rightarrow N_2^+ + e^-$ & $3.0 \cdot 10^{-22} T^{1.2} e^{\frac{80.0}{T^{1.0}}}$ & $m^3s^{-1}$ & \cite{popov_associative_2009, leonov_femtosecond_2012, volynets_n2_2018} \\ 
R189 & $2 N(P) \rightarrow N_2^+ + e^-$ & $5.796 \cdot 10^{-18} T^{0.1667}$ & $m^3s^{-1}$ & \cite{popov_associative_2009, leonov_femtosecond_2012, volynets_n2_2018} \\ 
R190 & $N(P) + O \rightarrow NO^+ + e^-$ & $5.796 \cdot 10^{-18} T^{0.1667}$ & $m^3s^{-1}$ & \cite{popov_associative_2009, leonov_femtosecond_2012, volynets_n2_2018} \\ 
R191 & $O_2 + e^- \rightarrow 2 O + e^-$ & $10^{-6} \cdot 10.0^{- \frac{\left(0.0005402 T_{e}^{1.178} + 1.777 \cdot 10^{-8} T_{e}^{2.0} + 8.4\right)}{\left(6.501 \cdot 10^{-5} T_{e}^{1.178} + 2.139 \cdot 10^{-9} T_{e}^{2.0}\right)^{1.0}}}$ & $m^3s^{-1}$ & \cite{poggieNumericalSimulationNanosecondpulse2012, kossyi_kinetic_1992} \\ 
R192 & $O_2 + e^- \rightarrow O + O(D) + e^-$ & $10^{-6} \cdot 10.0^{- \frac{\left(0.000511 T_{e}^{1.178} + 1.681 \cdot 10^{-8} T_{e}^{2.0} + 17.2\right)}{\left(6.501 \cdot 10^{-5} T_{e}^{1.178} + 2.139 \cdot 10^{-9} T_{e}^{2.0}\right)^{1.0}}}$ & $m^3s^{-1}$ & \cite{poggieNumericalSimulationNanosecondpulse2012, kossyi_kinetic_1992} \\ 
R193 & $2 N_2 \rightarrow 2 N + N_2$ & $\frac{2.019 \cdot 10^{-10} e^{- \frac{1.132 \cdot 10^{5}}{\left(T T_{v}\right)^{0.5}}}}{\left(T T_{v}\right)^{0.607}}$ & $m^3s^{-1}$ & \cite{park_rate_2008, niu_assessment_2018, kim_modification_2021} \\ 
R194 & $2 N + N_2 \rightarrow 2 N_2$ & $8.27 \cdot 10^{-46} e^{\frac{500.0 - 6.918 \cdot 10^{-7} T^{2.54}}{T^{1.0}}}$ & $m^6s^{-1}$ & \cite{park_rate_2008, niu_assessment_2018, kim_modification_2021, campbell_recombination_1997} \\ 
R195 & $N_2 + O_2 \rightarrow 2 N + O_2$ & $\frac{1.162 \cdot 10^{-8} e^{- \frac{1.132 \cdot 10^{5}}{\left(T T_{v}\right)^{0.5}}}}{\left(T T_{v}\right)^{0.8}}$ & $m^3s^{-1}$ & \cite{park_rate_2008, niu_assessment_2018, kim_modification_2021} \\ 
R196 & $N_2 + NO \rightarrow 2 N + NO$ & $\frac{1.162 \cdot 10^{-8} e^{- \frac{1.132 \cdot 10^{5}}{\left(T T_{v}\right)^{0.5}}}}{\left(T T_{v}\right)^{0.8}}$ & $m^3s^{-1}$ & \cite{park_rate_2008, niu_assessment_2018, kim_modification_2021} \\ 
R197 & $2 N + NO \rightarrow N_2 + NO$ & $8.27 \cdot 10^{-46} e^{\frac{500.0 - 6.918 \cdot 10^{-7} T^{2.54}}{T^{1.0}}}$ & $m^6s^{-1}$ & \cite{park_rate_2008, niu_assessment_2018, kim_modification_2021} \\ 
R198 & $N + N_2 \rightarrow 3 N$ & $\frac{5.963 \cdot 10^{-10} e^{- \frac{1.132 \cdot 10^{5}}{\left(T T_{v}\right)^{0.5}}}}{\left(T T_{v}\right)^{0.613}}$ & $m^3s^{-1}$ & \cite{park_rate_2008, niu_assessment_2018, kim_modification_2021} \\ 
R199 & $3 N \rightarrow N + N_2$ & $8.27 \cdot 10^{-46} e^{\frac{500.0 - 6.918 \cdot 10^{-7} T^{2.54}}{T^{1.0}}}$ & $m^6s^{-1}$ & \cite{park_rate_2008, niu_assessment_2018, kim_modification_2021} \\ 
R200 & $N_2 + O \rightarrow 2 N + O$ & $\frac{4.982 \cdot 10^{-8} e^{- \frac{1.132 \cdot 10^{5}}{\left(T T_{v}\right)^{0.5}}}}{\left(T T_{v}\right)^{0.8}}$ & $m^3s^{-1}$ & \cite{park_rate_2008, niu_assessment_2018, kim_modification_2021} \\ 
R201 & $2 N + O \rightarrow N_2 + O$ & $8.27 \cdot 10^{-46} e^{\frac{500.0 - 6.918 \cdot 10^{-7} T^{2.54}}{T^{1.0}}}$ & $m^6s^{-1}$ & \cite{park_rate_2008, niu_assessment_2018, kim_modification_2021} \\ 
R202 & $N_2 + N_2^+ \rightarrow 2 N + N_2^+$ & $\frac{2.019 \cdot 10^{-10} e^{- \frac{1.132 \cdot 10^{5}}{\left(T T_{v}\right)^{0.5}}}}{\left(T T_{v}\right)^{0.607}}$ & $m^3s^{-1}$ & \cite{park_rate_2008, niu_assessment_2018, kim_modification_2021} \\ 
R203 & $2 N + N_2^+ \rightarrow N_2 + N_2^+$ & $8.27 \cdot 10^{-46} e^{\frac{500.0 - 6.918 \cdot 10^{-7} T^{2.54}}{T^{1.0}}}$ & $m^6s^{-1}$ & \cite{park_rate_2008, niu_assessment_2018, kim_modification_2021} \\ 
R204 & $N_2 + O_2^+ \rightarrow 2 N + O_2^+$ & $\frac{1.162 \cdot 10^{-8} e^{- \frac{1.132 \cdot 10^{5}}{\left(T T_{v}\right)^{0.5}}}}{\left(T T_{v}\right)^{0.8}}$ & $m^3s^{-1}$ & \cite{park_rate_2008, niu_assessment_2018, kim_modification_2021} \\ 
R205 & $2 N + O_2^+ \rightarrow N_2 + O_2^+$ & $8.27 \cdot 10^{-46} e^{\frac{500.0 - 6.918 \cdot 10^{-7} T^{2.54}}{T^{1.0}}}$ & $m^6s^{-1}$ & \cite{park_rate_2008, niu_assessment_2018, kim_modification_2021} \\ 
R206 & $N_2 + NO^+ \rightarrow 2 N + NO^+$ & $\frac{1.162 \cdot 10^{-8} e^{- \frac{1.132 \cdot 10^{5}}{\left(T T_{v}\right)^{0.5}}}}{\left(T T_{v}\right)^{0.8}}$ & $m^3s^{-1}$ & \cite{park_rate_2008, niu_assessment_2018, kim_modification_2021} \\ 
R207 & $N_2 + N^+ \rightarrow 2 N + N^+$ & $\frac{5.963 \cdot 10^{-10} e^{- \frac{1.132 \cdot 10^{5}}{\left(T T_{v}\right)^{0.5}}}}{\left(T T_{v}\right)^{0.613}}$ & $m^3s^{-1}$ & \cite{park_rate_2008, niu_assessment_2018, kim_modification_2021} \\ 
R208 & $N_2 + O^+ \rightarrow 2 N + O^+$ & $\frac{4.982 \cdot 10^{-8} e^{- \frac{1.132 \cdot 10^{5}}{\left(T T_{v}\right)^{0.5}}}}{\left(T T_{v}\right)^{0.8}}$ & $m^3s^{-1}$ & \cite{park_rate_2008, niu_assessment_2018, kim_modification_2021} \\ 
R209 & $N_2 + O_2 \rightarrow N_2 + 2 O$ & $\frac{5.57 \cdot 10^{-15} e^{- \frac{5.95 \cdot 10^{4}}{T^{1.0}}}}{T^{0.2726}}$ & $m^3s^{-1}$ & \cite{park_rate_2008, niu_assessment_2018, kim_modification_2021} \\ 
R210 & $2 O_2 \rightarrow 2 O + O_2$ & $\frac{1.855 \cdot 10^{-5} e^{- \frac{5.95 \cdot 10^{4}}{T^{1.0}}}}{T^{2.585}}$ & $m^3s^{-1}$ & \cite{park_rate_2008, niu_assessment_2018, kim_modification_2021} \\ 
R211 & $NO + O_2 \rightarrow NO + 2 O$ & $\frac{5.57 \cdot 10^{-15} e^{- \frac{5.95 \cdot 10^{4}}{T^{1.0}}}}{T^{0.2726}}$ & $m^3s^{-1}$ & \cite{park_rate_2008, niu_assessment_2018, kim_modification_2021} \\ 
R212 & $NO + 2 O \rightarrow NO + O_2$ & $\frac{6.332 \cdot 10^{-53} e^{\frac{3310.0 T - 0.0001567 T^{3.0} - 5.237 \cdot 10^{5}}{T^{2.0}}}}{T^{0.2726} \left(T^{-1.0}\right)^{1.909}}$ & $m^6s^{-1}$ & \cite{park_rate_2008, niu_assessment_2018, kim_modification_2021} \\ 
R213 & $N + O_2 \rightarrow N + 2 O$ & $\frac{1.661 \cdot 10^{-8} e^{- \frac{5.95 \cdot 10^{4}}{T^{1.0}}}}{T^{1.5}}$ & $m^3s^{-1}$ & \cite{park_rate_2008, niu_assessment_2018, kim_modification_2021} \\ 
R214 & $O + O_2 \rightarrow 3 O$ & $\frac{4.982 \cdot 10^{-9} e^{- \frac{5.95 \cdot 10^{4}}{T^{1.0}}}}{T^{1.5}}$ & $m^3s^{-1}$ & \cite{park_rate_2008, niu_assessment_2018, kim_modification_2021} \\ 
R215 & $3 O \rightarrow O + O_2$ & $\frac{5.664 \cdot 10^{-47} e^{\frac{3310.0 T - 0.0001567 T^{3.0} - 5.237 \cdot 10^{5}}{T^{2.0}}}}{T^{1.5} \left(T^{-1.0}\right)^{1.909}}$ & $m^6s^{-1}$ & \cite{park_rate_2008, niu_assessment_2018, kim_modification_2021} \\ 
R216 & $N_2^+ + O_2 \rightarrow N_2^+ + 2 O$ & $\frac{5.57 \cdot 10^{-15} e^{- \frac{5.95 \cdot 10^{4}}{T^{1.0}}}}{T^{0.2726}}$ & $m^3s^{-1}$ & \cite{park_rate_2008, niu_assessment_2018, kim_modification_2021} \\ 
R217 & $O_2 + O_2^+ \rightarrow 2 O + O_2^+$ & $\frac{1.855 \cdot 10^{-5} e^{- \frac{5.95 \cdot 10^{4}}{T^{1.0}}}}{T^{2.585}}$ & $m^3s^{-1}$ & \cite{park_rate_2008, niu_assessment_2018, kim_modification_2021} \\ 
R218 & $NO^+ + O_2 \rightarrow NO^+ + 2 O$ & $\frac{5.57 \cdot 10^{-15} e^{- \frac{5.95 \cdot 10^{4}}{T^{1.0}}}}{T^{0.2726}}$ & $m^3s^{-1}$ & \cite{park_rate_2008, niu_assessment_2018, kim_modification_2021} \\ 
R219 & $N^+ + O_2 \rightarrow N^+ + 2 O$ & $\frac{1.661 \cdot 10^{-8} e^{- \frac{5.95 \cdot 10^{4}}{T^{1.0}}}}{T^{1.5}}$ & $m^3s^{-1}$ & \cite{park_rate_2008, niu_assessment_2018, kim_modification_2021} \\ 
R220 & $O_2 + O^+ \rightarrow 2 O + O^+$ & $\frac{4.982 \cdot 10^{-9} e^{- \frac{5.95 \cdot 10^{4}}{T^{1.0}}}}{T^{1.5}}$ & $m^3s^{-1}$ & \cite{park_rate_2008, niu_assessment_2018, kim_modification_2021} \\ 
R221 & $N_2 + NO \rightarrow N + N_2 + O$ & $2.408 \cdot 10^{-15} e^{- \frac{7.52 \cdot 10^{4}}{T^{1.0}}}$ & $m^3s^{-1}$ & \cite{park_rate_2008, niu_assessment_2018, kim_modification_2021} \\ 
R222 & $N + N_2 + O \rightarrow N_2 + NO$ & $\frac{5.76 \cdot 10^{-53} e^{\frac{3520.0 T - 0.0002093 T^{3.0} - 5.586 \cdot 10^{5}}{T^{2.0}}}}{\left(T^{-1.0}\right)^{2.028}}$ & $m^6s^{-1}$ & \cite{park_rate_2008, niu_assessment_2018, kim_modification_2021} \\ 
R223 & $NO + O_2 \rightarrow N + O + O_2$ & $2.408 \cdot 10^{-15} e^{- \frac{7.52 \cdot 10^{4}}{T^{1.0}}}$ & $m^3s^{-1}$ & \cite{park_rate_2008, niu_assessment_2018, kim_modification_2021} \\ 
R224 & $N + O + O_2 \rightarrow NO + O_2$ & $\frac{5.76 \cdot 10^{-53} e^{\frac{3520.0 T - 0.0002093 T^{3.0} - 5.586 \cdot 10^{5}}{T^{2.0}}}}{\left(T^{-1.0}\right)^{2.028}}$ & $m^6s^{-1}$ & \cite{park_rate_2008, niu_assessment_2018, kim_modification_2021} \\ 
R225 & $2 NO \rightarrow N + NO + O$ & $1.601 \cdot 10^{-15} e^{- \frac{7.52 \cdot 10^{4}}{T^{1.0}}}$ & $m^3s^{-1}$ & \cite{park_rate_2008, niu_assessment_2018, kim_modification_2021} \\ 
R226 & $N + NO + O \rightarrow 2 NO$ & $\frac{3.829 \cdot 10^{-53} e^{\frac{3520.0 T - 0.0002093 T^{3.0} - 5.586 \cdot 10^{5}}{T^{2.0}}}}{\left(T^{-1.0}\right)^{2.028}}$ & $m^6s^{-1}$ & \cite{park_rate_2008, niu_assessment_2018, kim_modification_2021} \\ 
R227 & $N + NO \rightarrow 2 N + O$ & $1.601 \cdot 10^{-15} e^{- \frac{7.52 \cdot 10^{4}}{T^{1.0}}}$ & $m^3s^{-1}$ & \cite{park_rate_2008, niu_assessment_2018, kim_modification_2021} \\ 
R228 & $2 N + O \rightarrow N + NO$ & $\frac{3.829 \cdot 10^{-53} e^{\frac{3520.0 T - 0.0002093 T^{3.0} - 5.586 \cdot 10^{5}}{T^{2.0}}}}{\left(T^{-1.0}\right)^{2.028}}$ & $m^6s^{-1}$ & \cite{park_rate_2008, niu_assessment_2018, kim_modification_2021} \\ 
R229 & $NO + O \rightarrow N + 2 O$ & $1.601 \cdot 10^{-15} e^{- \frac{7.52 \cdot 10^{4}}{T^{1.0}}}$ & $m^3s^{-1}$ & \cite{park_rate_2008, niu_assessment_2018, kim_modification_2021} \\ 
R230 & $N + 2 O \rightarrow NO + O$ & $\frac{3.829 \cdot 10^{-53} e^{\frac{3520.0 T - 0.0002093 T^{3.0} - 5.586 \cdot 10^{5}}{T^{2.0}}}}{\left(T^{-1.0}\right)^{2.028}}$ & $m^6s^{-1}$ & \cite{park_rate_2008, niu_assessment_2018, kim_modification_2021} \\ 
R231 & $N_2^+ + NO \rightarrow N + N_2^+ + O$ & $2.408 \cdot 10^{-15} e^{- \frac{7.52 \cdot 10^{4}}{T^{1.0}}}$ & $m^3s^{-1}$ & \cite{park_rate_2008, niu_assessment_2018, kim_modification_2021} \\ 
R232 & $NO + O_2^+ \rightarrow N + O + O_2^+$ & $2.408 \cdot 10^{-15} e^{- \frac{7.52 \cdot 10^{4}}{T^{1.0}}}$ & $m^3s^{-1}$ & \cite{park_rate_2008, niu_assessment_2018, kim_modification_2021} \\ 
R233 & $NO + NO^+ \rightarrow N + NO^+ + O$ & $1.601 \cdot 10^{-15} e^{- \frac{7.52 \cdot 10^{4}}{T^{1.0}}}$ & $m^3s^{-1}$ & \cite{park_rate_2008, niu_assessment_2018, kim_modification_2021} \\ 
R234 & $NO + N^+ \rightarrow N + N^+ + O$ & $1.601 \cdot 10^{-15} e^{- \frac{7.52 \cdot 10^{4}}{T^{1.0}}}$ & $m^3s^{-1}$ & \cite{park_rate_2008, niu_assessment_2018, kim_modification_2021} \\ 
R235 & $NO + O^+ \rightarrow N + O + O^+$ & $1.601 \cdot 10^{-15} e^{- \frac{7.52 \cdot 10^{4}}{T^{1.0}}}$ & $m^3s^{-1}$ & \cite{park_rate_2008, niu_assessment_2018, kim_modification_2021} \\ 
R236 & $N_2 + O_2 \rightarrow NO + NO^+ + e^-$ & $\frac{2.292 \cdot 10^{-10} e^{- \frac{1.41 \cdot 10^{5}}{T^{1.0}}}}{T^{1.84}}$ & $m^3s^{-1}$ & \cite{park_rate_2008, niu_assessment_2018, kim_modification_2021} \\ 
R237 & $2 N \rightarrow N_2^+ + e^-$ & $2.7 \cdot 10^{-17} e^{- \frac{6.74 \cdot 10^{4}}{T^{1.0}}}$ & $m^3s^{-1}$ & \cite{park_rate_2008, niu_assessment_2018, kim_modification_2021} \\ 
R238 & $2 O \rightarrow O_2^+ + e^-$ & $\frac{2.657 \cdot 10^{-13} e^{- \frac{8.08 \cdot 10^{4}}{T^{1.0}}}}{T^{0.98}}$ & $m^3s^{-1}$ & \cite{park_rate_2008, niu_assessment_2018, kim_modification_2021} \\ 
R239 & $N + O \rightarrow NO^+ + e^-$ & $1.499 \cdot 10^{-20} T^{0.5} e^{- \frac{3.24 \cdot 10^{4}}{T^{1.0}}}$ & $m^3s^{-1}$ & \cite{park_rate_2008, niu_assessment_2018, kim_modification_2021} \\ 
R240 & $N + e^- \rightarrow N^+ + 2 e^-$ & $\frac{182.7 e^{- \frac{1.69 \cdot 10^{5}}{T^{1.0}}}}{T^{3.14}}$ & $m^3s^{-1}$ & \cite{park_rate_2008, niu_assessment_2018, kim_modification_2021} \\ 
R241 & $O + e^- \rightarrow O^+ + 2 e^-$ & $\frac{59.78 e^{- \frac{1.58 \cdot 10^{5}}{T^{1.0}}}}{T^{2.91}}$ & $m^3s^{-1}$ & \cite{park_rate_2008, niu_assessment_2018, kim_modification_2021} \\ 


\bottomrule
\end{longtable}
\end{widetext}


\bibliography{ref_scitech_sagar,sciTech,refs_hwi,prelims,thesisZotero,plasma,g_scholar}

\end{document}